\documentclass[11pt,a4paper]{article}
\usepackage{float}
\usepackage{multirow}
\usepackage[cp1252,utf8]{inputenc}
\usepackage{hyperref}
\usepackage{graphicx}
\usepackage{wrapfig}
\usepackage{amsmath}
\usepackage{xparse}
\usepackage{amsfonts}
\usepackage{amssymb}
\usepackage{relsize} 
\usepackage[top=1in, bottom=2cm, left=1.5cm, right=1.5cm]{geometry}
\usepackage{authblk}
\usepackage{ulem}
\usepackage{physics}
\usepackage{xcolor}
\NewDocumentCommand{\tens}{t_}
{%
	\IfBooleanTF{#1}
	{\tensop}
	{\otimes}%
}
\NewDocumentCommand{\tensop}{m}
{%
	\mathbin{\mathop{\otimes}\displaylimits_{#1}}%
}
\usepackage{changepage} 

\usepackage{afterpage}

\usepackage{placeins}

\usepackage{wrapfig}
\usepackage{cite} 
\usepackage[nottoc,notlof,notlot]{tocbibind} 

\title{\bf Holographic Brownian dynamics of a heavy particle in a boosted thermal plasma background}
\author[a]{\bf  Anirban Roy Chowdhury \thanks{iamanirban@bose.res.in}}
\author[b]{\bf  Ashis Saha \thanks{ashis.hepth@gmail.com}}
\author[c]{\bf Sunandan Gangopadhyay \thanks{sunandan.gangopadhyay@bose.res.in}}

\affil[a,c]{\textit{Department of Astrophysics and High Energy Physics\\}
\textit{S.N.~Bose National Centre for Basic Sciences,}
\textit{JD Block, Sector-III, Salt Lake, Kolkata 700106, India}}
\affil[b]{\textit{Department of Physics, School of Basic and Applied Sciences,Adamas University, Adamas Knowledge City, Kolkata,  700126, India}}

\date{}

\begin{document}
	\maketitle
	\begin{abstract}
		\noindent  In this work, we have performed a detailed holographic analysis of the stochastic dynamics of a heavy particle propagating through a strongly coupled plasma moving with a constant velocity along a fixed spatial direction. To model this scenario within the framework of the AdS/CFT correspondence, we consider a boosted AdS black brane geometry in the bulk. The boost corresponds to the uniform motion of the plasma on the boundary field theory side. The presence of this boost introduces a preferred direction, leading to an anisotropic environment in which the behavior of the Brownian particle differs depending on its direction of motion. Consequently, we examine two distinct cases, namely,Brownian motion parallel to the direction of the boost and motion perpendicular to it. In this work we have computed the diffusion coefficient for both along the boost and perpendicular to the boost directions. We have obtained the diffusion coefficient by following the two different approaches in both the cases. These complementary approaches yield consistent results, thereby reinforcing the reliability of the computations carried out. Additionally, we verify the fluctuation-dissipation theorem within this anisotropic setup, confirming its validity in both longitudinal and transverse to the direction of boost. Our findings provide deeper insight into the non-equilibrium transport properties of strongly coupled plasma and further elucidate the holographic description of Brownian motion in anisotropic backgrounds. Finally, we proceed to holographically compute the Butterfly velocity by using the entanglement wedge subregion duality and express the diffusion coefficients in terms of the chaotic observables.
		\end{abstract}
	\section{Introduction}
	\noindent One of the central questions in statistical mechanics is to understand the emergence of macroscopic dissipation and the process of thermalization from an underlying microscopic perspective. In conventional treatments, particularly in the thermodynamic or hydrodynamic limit these macroscopic phenomena are typically attributed to the collective effect of microscopic collisions among the constituents of the system. This viewpoint is fundamentally rooted in the concept of Brownian motion. In 1827, the botanist Robert Brown observed that pollen grains suspended in water exhibit a persistent and erratic motion under a microscope, a phenomenon later attributed to random collisions with thermally agitated fluid molecules \cite{Brown01091828,Mori:1965oqj}. In recent days, it is well established that Brownian motion exemplifies the influence of microscopic degrees of freedom on macroscopic behavior. Any particle immersed in a thermal medium experiences such stochastic motion, whether it is a small pendulum suspended in a dilute gas \cite{Uhlenbeck:1930zz}, or a heavy quark traversing a quark-gluon plasma. This universal behavior strongly suggests that the interaction between macroscopic objects \cite{Chandrasekhar:1943ws,Dunkel:2008ngc} and their microscopic environments underlies both dissipation and the approach to thermal equilibrium \cite{Feynman:1963fq,Hu:1991di,Hu:1993vs,Wu:2005jr,Hsiang:2011by}.\\
	On the other hand, it is shown that one can study the dynamics of non-Abelian quark-gluon plasm at finite temperature with the help of AdS/CFT duality \cite{Maldacena:1997re,Gubser:1998bc,Witten:1998qj,Aharony:1999ti}. This duality provides us a systematic way to study the dynamics of quark gluon plasma (QGP) \cite{Collins:1974ky,Shuryak:2014zxa,BRAHMS:2004adc,PHOBOS:2004zne,Bluhm:2020mpc,Gubser:2007zz,Mateos:2007ay,Son:2007zz} by considering an appropriate gravity theory in the bulk. In recent studies it is also shown that this AdS/CFT duality provides a suitable tool to study the hydrodynamic regimes of the quark-gluon plasma \cite{Panero:2009tv,Policastro:2001yc,Kim:2007xi,Casalderrey-Solana:2011dxg,Arefeva:2020vae,Colangelo:2010pe,Li:2012ay,He:2020fdi,Cao:2020ryx,Cao:2020ske}. This correspondence enables a detailed and quantitative investigation of strongly coupled plasmas through their dual gravitational descriptions in the bulk, and conversely, offers insights into gravitational dynamics via boundary field theories \cite{Braga:2019xwl,Braga:2020myi,Rodrigues:2020ndy,Chen:2020ath,Ballon-Bayona:2020xtf,Mes:2020vgy,Giataganas:2013zaa}. Given this powerful duality, it is natural to ask whether Brownian motion, a quintessential example of stochastic behavior arising from microscopic interactions, admits a holographic description. Exploring this question not only deepens our understanding of non-equilibrium processes in strongly coupled systems but also serves as a step toward uncovering the microscopic foundations of thermodynamics and hydrodynamics within the holographic framework. The study of linear response in the context of Brownian motion via holographic models was initiated in \cite{deBoer:2008gu,Atmaja:2010uu,Gubser:2006bz,Casalderrey-Solana:2006fio,Gubser:2006nz}, where a novel setup was proposed involving a stretched string extending from the horizon of an AdS black hole to a probe brane located near the boundary of the spacetime. The presence of the black hole horizon plays a crucial role, as it enables the definition of a Hawking temperature, allowing key physical quantities to be expressed as functions of temperature. In this framework, the endpoint of the string on the probe brane is interpreted as a heavy probe particle immersed in a thermal medium. By analyzing the fluctuations of the string near the boundary, one can infer the stochastic motion of the particle and compute the diffusion coefficient through the system's admittance. Furthermore, the model captures the characteristic time evolution of the mean square displacement, successfully reproducing both the short-time ballistic regime and the long-time diffusive behavior, hallmarks of Brownian motion \cite{Chakrabortty:2013kra,Sadeghi:2013jja,Banerjee:2013rca,Banerjee:2015vmo,Chakrabarty:2019aeu,Zhou:2024khk,Bu:2022oty,Banerjee:2017set,Jahnke:2016gkb,Chakraborty:2014kfa,Casalderrey-Solana:2007ahi}.
	This holographic setup has also been extended to explore quantum critical behavior, particularly in backgrounds exhibiting Lifshitz and hyperscaling-violating Lifshitz symmetries \cite{Yeh:2013mca,Edalati:2012tc}. These geometries are employed to model non-relativistic and scale-invariant systems, allowing for the identification and characterization of quantum critical points within a strongly coupled regime. Furthermore, some recent interesting works in this direction can be found in \cite{Rajagopal:2025ukd,Zhou:2024pbb,Zhou:2024oeg,Caldeira:2022gfo,Sivakumar:2022viy,Bu:2021jlp,Caldeira:2021izy,Caldeira:2020rir,Caldeira:2020sot,Cartas-Fuentevilla:2020ntr,Kundu:2019ull,Giataganas:2018rbq,Giataganas:2018ekx,Yeh:2015cra,TangarifeGarcia:2014uxa}.\\
	In recent years, the study of chaotic systems, particularly in the context of quantum many-body dynamics and black hole physics has uncovered a variety of rich and surprising physical phenomena, such as fast scrambling, operator growth, and universal bounds on chaos. One can study a chaotic system by computing the Lyapunov exponent associated with the system. In the context of classical system this Lyapunov exponent ($\lambda_{L}$) can be defined as
	\begin{eqnarray}
		\lambda_{L}\sim \frac{1}{t}\log(\frac{\delta x(t)}{\delta x(0)})
	\end{eqnarray}
	where $\delta x(t)$ represents a small deviation from the classical phase space trajectory, induced by an infinitesimal change in the initial conditions. In chaotic systems, such perturbations typically grow exponentially with time. The nature of a dynamical system, whether it is chaotic, stable, or marginally stable, can be characterized by the value of Lyapunov exponent, a positive Lyapunov exponent indicates chaos (sensitive dependence on initial conditions), a zero value suggests neutral stability (as in integrable systems), and a negative exponent corresponds to stable, non-chaotic behavior where perturbations decay over time. However, the study of chaos in the quantum regime is highly nontrivial, primarily due to the absence of classical trajectories. In recent studies it was shown that one can study quantum chaos by computing the double commutator of two hermitian operators $\hat{V}(0)$ and $\hat{W}(x,t)$. The double commutator between these two operators is defined as \cite{Larkin1969QuasiclassicalMI}
	\begin{eqnarray}
		\mathcal{C}&=&-\langle [\hat{W}(x,t),\hat{V}(0)]^2 \rangle_{\beta}\nonumber\\
		&\approx&\langle \hat{W}(x,t)\hat{V}(0)\hat{W}(x,t)\hat{V}(0)\rangle_{\beta}
	\end{eqnarray}
	The quantity \(\mathcal{C}(x, t) \) quantifies the sensitivity of the later-time observable \( \hat{W}(x, t) \) to an earlier perturbation induced by \( \hat{V}(0) \), capturing the hallmark of chaotic dynamics in quantum systems. This above quantity is also useful to study the rate at which information can spread between two spacelike points. This leads to the emergence of a characteristic speed, often referred to as the butterfly velocity, which quantifies the rate at which the influence of a local perturbation spreads through the system.
	However, it is also shown that, for large $N$ strongly coupled field theories, the above commutator has the following form \cite{Shenker:2013pqa,Shenker:2014cwa}
	\begin{eqnarray}
		\mathcal{C}(t)\approx\lambda_{L}\left(t-t_{*}-\frac{|x|}{v_{B}}\right) 
	\end{eqnarray} 
	where $\lambda_{L}$ is the Lyapunov exponent, obeying the MSS bound, that is, $\lambda_{L}\le\frac{2\pi}{\beta}$ ($\beta$ is inverse of Hawking temperature)\cite{Stanford:2015owe,Maldacena:2015waa}, $t_{*}$ is the scrambling time at which $\mathcal{C}(t)\sim 1$, $|x|$ represents the separation between the two points and $v_{B}$ denotes the butterfly velocity which indicates the growth of a given perturbation $\hat{V}(0)$. In \cite{Blake_2016,Blake_2016A}, it was shown that there exists an universal regime in which charge diffusion constant and energy diffusion constant (for a theory with holographic dual) can be expressed in terms of the observables of the quantum chaos, that is, Lyapunov exponent $\lambda_{L}$ and butterfly velocity $v_B$. These relations in turn reveal the subtle deep connection existing between the microscopic transport coefficients and the chaotic observables. Motivated by this observation, one can look for same chaotic description for the momentum diffusion constant for the heavy particle in this context. In order to do this we first need to compute the butterfly velocity for the boosted black brane geometry. There are different ways to compute butterfly velocity in the context of AdS/CFT correspondence \cite{Plamadeala:2018vsr,Das:2022jrr,Gu:2016hoy,Lin:2018tce,Khetrapal:2022dzy,Shenker:2014cwa,Roberts:2014isa,Maldacena:2015waa}. In this work we have followed the recently developed entanglement wedge sub-region duality to compute the butterfly velocity \cite{Mezei:2019zyt,Fischler:2018kwt,Dong:2022ucb}. 
	In this present work we have extended these studies in the boosted black bane background setup. This boosted black brane geometry in the bulk describes a strongly coupled plasma at finite temperature, moving with a uniform velocity along a particular direction. We have calculated the mean square displacement for both the fermionic and bosonic case by considering the Fermi-Dirac and Bose-Einstein distribution functions respectively. Then we take the ballistic and diffusive limit to the mean square displacement. It is to be noted that in the diffusive limit we have obtained the expression of diffusion coefficient. On the other hand we have also obtained the diffusion coefficient by computing the admittance. Then we moved on to check the fluctuation-dissipation theorem. We have carried out our study in both the parallel and perpendicular directions relative to the direction of boost. In this work, we investigate the behavior of the diffusion coefficient in the presence of a Lorentz boost. Our analysis reveals that the diffusion coefficient exhibits a decreasing trend with increasing boost parameter, highlighting a suppression of diffusion. Specifically, we find that the diffusion coefficient along the direction of the boost is significantly smaller than its counterpart in the transverse direction. This anisotropy reflects the inherent directional dependence introduced by the boost and has important implications for the dynamics of strongly coupled systems in non-static backgrounds.
	\section{Brief introduction to Brownian motion}
	\noindent In this section we would like to briefly discuss about the Brownian motion and Langevin dynamics. Langevin equation describes the Brownian motion of a non-relativistic particle. Now let us consider a large particle of mass $m$ is immersed in a viscous fluid of much smaller particles. The equation of motion for this particle can be written as \cite{article,Kubo:1966fyg,Einstein:1905rjn,10.1143/PTP.33.423,10.1143/PTP.34.399}
	\begin{eqnarray}\label{eq1}
		m\frac{dv(t)}{dt}=F(t)
	\end{eqnarray}
	where $v(t)$ is the instantaneous velocity of the particle and $F(t)$ represents the total instantaneous force acting on the particle. This total instantaneous force again consists of two friction forces which arises due to the interaction between the Brownian particle and the fluid medium, on the other hand there is a random force due to the density fluctuation in the fluid. Therefore, the total instantaneous force on the Brownian particle can be written as 
	\begin{eqnarray}\label{eq2}
		F(t)=-\gamma v(t)+\xi(t)
	\end{eqnarray}
	where $\gamma$ is the friction coefficient which depends on the size ($a$) of the Brownian particle and the viscosity $(\eta)$ of the fluid medium. On the other hand $\xi (t)$ represents the random force acting on the Brownian particle. This force is supposed to vary rapidly over time. The effect of this random force can summarized by its first and second moments, which are given
	\begin{eqnarray}
		\langle\xi(t)\rangle_{\xi}=0~~;~~\langle\xi(t)\xi(t^{\prime})\rangle_{\xi}=g\delta(t-t^{\prime})~~
	\end{eqnarray}  
	where $\langle...\rangle_{\xi}$ represents the average with respect to the appropriate probability distribution function of the random variable $\xi(t)$ and $g$ is a constant which represents the strength of the random force.\\
	Now substituting eq.\eqref{eq1} in eq.\eqref{eq2}, we have following equation of motion for the Brownian particle
	\begin{eqnarray}
		m\frac{dv(t)}{dt}=F(t)=-\gamma v(t)+\xi(t)~.
	\end{eqnarray}
	Upon solving the above equation for $v(t)$ and using the equipartition of energy, one can find the following relation between the $g$ and $\gamma$
	\begin{eqnarray}
		g=2\gamma k_{B}T
	\end{eqnarray}
	where $k_{B}$ is the Boltzmann constant and $T$ is the temperature of the surrounding medium. It is to be noted that the above relation is the simplest example of the fluctuation-dissipation theorem and arises due to the
	fact that frictional and random forces are of the same origin.\\
	Furthermore, solving the differential equation of displacement $x(t)$ and assuming the equipartition of energy, we can obtain the mean square displacement $s^{2}(t)$. Therefore, the expression of mean square displacement reads \cite{PhysRev.36.823,RevModPhys.17.323}
	\begin{eqnarray}
		\langle s^{2}(t)\rangle=\langle[x(t)-x(0)]^2\rangle\sim\begin{cases}
			\frac{k_B T}{m}t^2 ~~~~\text{for} ~~t\gamma<<1\\
			\frac{k_{B}T}{\gamma m} t ~~~ \text{for} ~~t\gamma>>1
		\end{cases}
	\end{eqnarray}
	where the time domains $t\gamma<<1$ and $t\gamma>>1$ are referred to the ballistic time domain and diffusive time domain. One can identify the diffusion coefficient from the expression of mean square displacement in the diffusive time domain. The expression of the diffusion coefficient reads
	\begin{eqnarray}
		D=\frac{k_B T}{\gamma m}~.
	\end{eqnarray} 
	The above relation is known as the Sutherland-
	Einstein relation \cite{PhysRev.36.823,RevModPhys.17.323}.
	\section{Boosted black brane geometry in $AdS_{d+1}$}
	In this section we will briefly discuss about the boosted black brane geometry. The ($d+1$)- dimensional gravitational background dual to large $N$, strongly coupled uniformly boosted plasma at a finite temperature in the $d$-dimensional boundary theory  can be described as boosted
	AdS-Schwarzschild black hole spacetime. The metric for this boosted black brane geometry is given by \cite{Bhatta:2019eog,Blanco:2013joa}
	\begin{eqnarray}\label{geo}
		ds^2 &=& \frac{r^2}{R^2} \left[ -dt^2 + dy^2 + \gamma^2 \left( \frac{r_H^d}{r^d} \right) (dt + v\,dy)^2 + dx_{d-2}^{\,2} \right]\nonumber\\ &+& \frac{R^2}{r^2} \frac{dr^2}{1 - \left( \frac{r_H}{r} \right)^d}~
	\end{eqnarray}
	where $\gamma=\frac{1}{\sqrt{1-v^2}}$ and in the natural unit, $v$ is dimensionless, which is bounded $0<v<1$. In the above metric we have assumed that the boost is along the $y$-direction. It is to be noted that the $r$ coordinate represents the bulk direction, therefore, the boundary is located at $r=\infty$ and $r=r_{H}$ denotes the location of the black hole horizon and $R$ is the AdS radius. The explicit form of different metric components reads \cite{Bhatta:2019eog,Blanco:2013joa}
	\begin{eqnarray}\label{matricc}
		g_{tt}&=-&\frac{r^2}{R^2}\left[1-\gamma^2\left(\frac{r_{h}}{r}\right)^d\right]~;~
		g_{rr}=\frac{R^2}{r^2}\frac{1}{1-\left(\frac{r_{h}}{r}\right)^d}\nonumber\\~g_{yy}&=&\frac{r^2}{R^2}\left(1+\gamma^2v^2\left(\frac{r_{h}}{r}\right)^d\right)\nonumber\\
		g_{ty}&=&v\gamma^2\left(\frac{r^2}{R^2}\right)\left(\frac{r_{h}}{r}\right)^d~;~
		g_{xx}=\frac{r^2}{R^2}~.
	\end{eqnarray}
	The Hawking temperature associated with the boosted black brane geometry is given by 
	\begin{eqnarray}\label{T}
		T=\frac{d}{4\pi R^2}\frac{r_{H}}{\gamma}~.
	\end{eqnarray}
	Note that, in the limit $v\rightarrow 0$, we recover the usual AdS-Schwarzschild black hole metric which describes a strongly coupled plasma at a finite temperature at rest. \\
	\section{Brownian motion along the boost}
	\noindent In this section we carry out our analysis by considering the fact that the heavy Brownian particle is moving through the  quark gluon plasma in the direction of the boost. One can analyze this problem from the bulk perspective by considering the fluctuation of an open string along the direction of the boost in the boosted black hole background (given in eq.\eqref{geo}).  To proceed further, we consider the Nambu-Goto (NG) action which describes the dynamics of the fluctuating string along the direction of boost. The NG action in this scenario reads
	\begin{eqnarray}\label{NG}
		S_{NG}^{para}=-\frac{1}{2\pi \alpha^\prime}\int d\tau d\sigma \sqrt{-\mathrm{det}\gamma_{ab}^{para}}
	\end{eqnarray}
	where $\tau$ and $\sigma$ represents the time and spatial coordinates on the string world sheet and $\gamma_{ab}^{para}$ is the two dimensional induced metric on the string world sheet. The general expression of the induced metric $\gamma_{ab}^{para}$ reads 
	\begin{eqnarray}
		\gamma_{ab}^{para}=g_{\mu\nu}\partial_{a}X^{\mu}\partial_{b}X^{\nu}~~;~~a,b=0,1~~; ~~ \mu,\nu=0,..., d+1
	\end{eqnarray}
	where $g_{\mu\nu}$ is the background bulk metric. To proceed further we consider the usual static gauge \cite{Giataganas:2018ekx,PhysRevD.87.046001,deBoer:2008gu} which implies that the time coordinate on the worldsheet is identified with the temporal coordinate of the bulk geometry and the spatial coordinate on the worldsheet is identified with the radial coordinate of the bulk metric. Therefore, $\tau=t$ and $\sigma=r$. The string fluctuation is parametrized by $Y(r,t)$.  Keeping these in mind, we now obtain the components of the two dimensional induced metric $\gamma_{ab}^{para}$. The components of $\gamma_{ab}^{para}$ read
	\begin{eqnarray}
		\gamma_{tt}^{para}&=&g_{tt}+g_{ty}\partial_{t}Y(t,r)+g_{yy}(\partial_{t}Y(r,t))^2\nonumber\\
		\gamma_{tr}^{para}&=& \gamma_{rt}^{para}=g_{yy}\partial_{t}Y(r,t)\partial_{r}Y(r,t)+g_{ty}\partial_{r}Y(t,r)\nonumber\\
		\gamma_{rr}^{para}&=&g_{rr}+g_{yy}(\partial_{r}Y(r,t))^2~.
	\end{eqnarray}
	The determinant of the induced metric can therefore be written as 
	\begin{eqnarray}
		\mathrm{det}(\gamma_{ab}^{para})&=&\gamma_{tt}^{para}\gamma_{rr}^{para}-(\gamma_{rt}^{para})^2\nonumber\\
		&\approx& g_{tt}g_{rr}+(g_{yy}g_{tt}-g_{ty}^2)\left(\partial_{r}Y(r,t)\right)^2+g_{yy}g_{rr}\left(\partial_{t}Y(r,t)\right)^2
	\end{eqnarray}
	In the above expression we have kept terms upto quadratic order, that is keeping  terms up to ($\partial_{t}Y)^2$ and $(\partial_{r}Y)^2$.
	Now substituting the above result in eq.\eqref{NG}, the Nambu-Goto action can be written as\footnote{Here we have used the fact that $g_{tt}=-|g_{tt}|$} 
		\begin{eqnarray}
			S_{NG}^{para}&=&-\frac{1}{2\pi\alpha^\prime}\int dt dr~\sqrt{-(g_{tt}g_{rr}+(g_{yy}g_{tt}-g_{ty}^2)\left(\partial_{r}Y(r,t)\right)^2+g_{yy}g_{rr}\left(\partial_{t}Y(r,t)\right)^2)}\nonumber\\
			&\approx&-\frac{1}{2\pi\alpha^\prime}\int dt dr\sqrt{|g_{tt}|g_{rr}}\left[1-\frac{\left(\partial_{r}Y(r,t)\right)^2}{2}\frac{(g_{tt}g_{yy}-g_{ty}^2)}{|g_{tt}|g{rr}}-\frac{\left(\partial_{t}Y(r,t)\right)^2}{2}\frac{g_{yy}}{|g_{tt}|}\right]~.\nonumber\\
		\end{eqnarray}
	To get the above result we make a binomial expansion and keep terms upto quadratic order in $\left(\partial_{r}Y(r,t)\right)$ and $\left(\partial_{t}Y(r,t)\right)$. Upon further simplification the result can be written as
\begin{eqnarray}\label{sng}
S_{NG}^{para}&=&S_{NG}^{(0)}+~\frac{1}{4\pi \alpha ^\prime}\int dt dr \Bigg[\frac{\left(\partial_{r}Y(r,t)\right)^2}{2}\frac{(g_{tt}g_{yy}-g_{ty}^2)}{\sqrt{|g_{tt}|g{rr}}}+\frac{\left(\partial_{t}Y(r,t)\right)^2}{2}\frac{g_{yy}\sqrt{g_{rr}}}{\sqrt{|g_{tt}|}}\Bigg]~.
\end{eqnarray}
	The variation of the above action leads to the following equation of motion
	\begin{eqnarray}\label{stf}
		\partial_{r}\Bigg[\frac{g_{tt}g_{yy}-g_{ty}^2}{\sqrt{|g_{tt}|g_{rr}}}\partial_{r}Y\Bigg]+\partial_{t}\Bigg[g_{yy}\sqrt{\frac{g_{rr}}{|g_{tt}|}}~\partial_{t} Y\Bigg]=0~.
	\end{eqnarray}
	It is to be noted that the first term in eq.\eqref{sng} does not contain any term involving $\partial_{r}Y$ and $\partial_t Y$ which indicates that this term does not have any contribution to the equation of motion. Only the second term in  eq.\eqref{sng} contributes to the equation of motion describing the string fluctuation. Furthermore, to get the above equation we have used the Neumann boundary condition instead of the Dirichlet boundary condition. The boundary condition in this scenario states that the spatial derivative of $Y$ should vanish at the boundary. It is given by
	\begin{eqnarray}\label{nbc}
		\frac{g_{tt}g_{yy}-g_{ty}^2}{\sqrt{|g_{tt}|g_{rr}}}\partial_{r}Y(r,t)|_{\mathrm{boundary}}=0~.
	\end{eqnarray}
	Now taking the ansatz $Y(r,t)=h_{\omega}(r)e^{i \omega t}$, one can recast the above equation in the following form
	\begin{eqnarray}\label{Y}
		\partial_{r}\Bigg[\frac{g_{ty}^2-g_{tt}g_{yy}}{\sqrt{|g_{tt}|g_{rr}}}\partial_{r}h_{\omega}\Bigg]+\omega^2g_{yy}\sqrt{\frac{g_{rr}}{|g_{tt}|}}h_{\omega}=0~.
	\end{eqnarray}
	To proceed further we now rewrite the above equation by substituting the expressions of the metric components appearing in the above equation by using eq.\eqref{matricc}. This gives
		\begin{eqnarray}\label{EQ1}
			\partial_{r}\Bigg[\frac{\left(\frac{r}{R}\right)^4 f^{\frac{3}{2}}(r)}{\left(1-\gamma^2\left(\frac{r_{h}}{r}\right)^d\right)^{\frac{1}{2}}}~~\partial_{r}h_{\omega}\Bigg]+\omega^2\frac{1+v^2\gamma^2\left(\frac{r_{h}}{r}\right)^d}{\sqrt{f(r)}\left(1-\gamma^2\left(\frac{r_{h}}{r}\right)^d\right)^{\frac{1}{2}}}~h_{\omega}=0~.
		\end{eqnarray}
	It is to be noted that for a generic non-diagonal spacetime geometry, it is not possible to solve the above equation analytically.
	\subsection{General solution}
	\noindent It is clear that it is not possible to find an exact analytical solution of eq.\eqref{Y}(or eq.\eqref{EQ1}). Therefore, we need to apply the standard patching method to obtain an approximate analytical solution. We need to solve this differential equation by considering the following cases:
	(a) near horizon limit (IR);
	(b) hydrodynamic limit, that is, $\omega\to 0$;
	(c) far from the horizon limit (UV).
	\subsubsection{Solution in the near horizon region}
	\noindent First, we consider the near horizon region which is also referred as the IR region. The solution of eq.\eqref{EQ1} in this regime is denoted by $h_{\omega}^{A}(r)$. To proceed further we need to recast the above equation in terms of the tortoise coordinate. The tortoise corrdinate for the bulk geometry or the boosted black brane geometry is given by
	\begin{eqnarray}
		\frac{dr_{*}}{dr}=\frac{\gamma R^2}{r^2 f(r)}~.
	\end{eqnarray}
	A detail derivation of the above expression in given in Appendix-II. It should be noted that the above expression for the tortoise coordinate differs from that of the standard Schwarzschild geometry only by an overall factor $\gamma$. In the limit $\gamma \to 1$, one recovers the usual Schwarzschild result.
		\noindent Keeping this above result in mind we can recast  eq.\eqref{EQ1} in the following form
		\begin{eqnarray}\label{EQ2}
			\partial_{r_*}\Bigg[\frac{\left(\frac{r}{R}\right)^2 f^{\frac{1}{2}}(r)}{\left(1-\gamma^2\left(\frac{r_{h}}{r}\right)^d\right)^{\frac{1}{2}}}~~\partial_{r_*}h_{\omega}\Bigg]+\frac{\omega^2}{\gamma^2}\frac{\left(\frac{r}{R}\right)^2\left(1+v^2\gamma^2\left(\frac{r_{h}}{r}\right)^d\right)\sqrt{f(r)}}{\left(1-\gamma^2\left(\frac{r_{h}}{r}\right)^d\right)^{\frac{1}{2}}}~h_{\omega}=0~.
		\end{eqnarray}
	For simplicity of the calculation we can write down the above equation as 
		\begin{eqnarray}\label{EQ3}
			\frac{d}{dr_*}\left[\alpha(r)\frac{dh_{\omega}}{dr_{*}}\right]+\frac{\omega^2}{\gamma^2}\delta(r)h_{\omega}=0
		\end{eqnarray}
		where the expression of $\alpha(r)$ and $\delta(r)$ are given by
		\begin{eqnarray}
			\alpha(r)&=&\frac{\left(\frac{r}{R}\right)^2 f^{\frac{1}{2}}(r)}{\left(1-\gamma^2\left(\frac{r_{h}}{r}\right)^d\right)^{\frac{1}{2}}}~~;~~
			\delta (r)=\frac{\left(\frac{r}{R}\right)^2\left(1+v^2\gamma^2\left(\frac{r_{h}}{r}\right)^d\right)\sqrt{f(r)}}{\left(1-\gamma^2\left(\frac{r_{h}}{r}\right)^d\right)^{\frac{1}{2}}}~~.
		\end{eqnarray}
	One can further simplify the eq.(\ref{EQ3}) and get 
	\begin{eqnarray}
		\frac{d^2h_{\omega}}{dr_*^2}+\frac{d\log(\alpha(r))}{dr_*}\frac{d h_{\omega}}{dr_*}+\frac{\omega^2}{\gamma^2}\frac{\delta(r)}{\alpha(r)}h_{\omega}=0~.
	\end{eqnarray}
	To proceed further we take the following ansatz of $h_{\omega}$, $h_{\omega}=e^{-\frac{A(r_{*})}{2}}\psi_{\omega}(r)$. Upon substituting the mentioned ansatz in the above equation yields
	\begin{eqnarray}\label{EQ4}
		\frac{d^2\psi}{dr_*^2}+\left[\frac{\omega^2}{\gamma^2}\frac{\delta(r)}{\alpha(r)}-V(r)\right]=0.
	\end{eqnarray}
	It should be noted that there is no term involving the first derivative of $\psi_{\omega}$(that is, there is no term with $\frac{d\psi}{dr_*}$). This is becuase of the following indentification
	\begin{eqnarray}
		A(r_{*})=\log(\alpha(r))~.
	\end{eqnarray}
	This is known as the Hills form\cite{magnus2004hill}. Kepping the indentification in mind the expression of $V(r)$ and $h_{\omega}$ can be written as 
	\begin{eqnarray}
		V(r)&=&\frac{1}{2}\frac{r^2 f(r)}{\gamma R^2}\frac{d}{dr}\left(\frac{r^2f(r)}{\gamma R^2}\frac{d \log(\alpha(r))}{dr}\right)+\frac{1}{4}\frac{r^4f^{2}(r)}{\gamma^2 r^4}\left(\frac{d \log(\alpha)}{dr}\right)^2\\
		h_{\omega}(r)&=&\frac{\psi_{\omega}(r)}{\sqrt{\alpha(r)}}
	\end{eqnarray}
	The above expression of $V(r)$ suggest that in the near horizon limit ($r\rightarrow \tilde{r}_{h}=r_{h}(1+2\epsilon)$\cite{deBoer:2008gu}), $V(r)$ should vanish. Therefore to proceed further we assume that $V(r)\approx0$ in this doamin. Keeping this fact in mind the equation of $\psi_{\omega}$ given in eq.\eqref{EQ4} can be written as 
	\begin{eqnarray}
		\frac{d^2\psi}{dr_*^2}+\frac{\omega^2}{\gamma^2}g(\tilde{r}_{h})\psi_{\omega}=0
	\end{eqnarray}
	where the expression of $g(\tilde{r}_{h})$ is given as, $g(\tilde{r}_{h})=1+v^2\gamma^2\left(\frac{r_{h}}{\tilde{r}_{h}}\right)^d$.
	A comprehensive derivation of the preceding equation can be found in \cite{Caldeira:2020rir,Caldeira:2020sot,Caldeira:2021izy}. 
	Now the general solution of the above equation can be written as 
	\begin{eqnarray}
		\psi(r)=A_{1}e^{-i \frac{\omega}{\gamma}\sqrt{g(\tilde{r}_{h})} r_{*}}+A_{2}e^{i \frac{\omega}{\gamma}\sqrt{g(\tilde{r}_{h})} r_{*}}
	\end{eqnarray}
	where $A_{1}$ and $A_{2}$ are arbitrary constants. To proceed further we only consider the ingoing solution, so we set $A_2=0$. This results in the following expression of $h_{\omega}^{A}(r)$
	\begin{eqnarray}
		h_{\omega}^{A}(r)&=&\frac{A_{1}e^{-i \frac{\omega}{\gamma}\sqrt{g(\tilde{r}_{h})} r_{*}}}{\sqrt{\alpha(\tilde{r}_{h})}}
		\approx \frac{A_{1}\left(1-i\frac{\omega}{\gamma}\sqrt{g(\tilde{r}_{h})}r_{*}\right)}{\sqrt{\alpha(\tilde{r}_{h})}}~.
	\end{eqnarray}
	In the second line of the above result, we have used the small frequency approximation. Before proceeding further, it is worth commenting on our choice of taking the near-horizon limit 
	$r \to \tilde r_h$ rather than $r \to r_h$. The above equation shows that taking the limit 
	$r \to r_h$ leads to a divergence in the solution. This divergence originates from the fact that 
	$\alpha(r) \propto f(r)$, where $f(r)$ is the blackening function that vanishes at the horizon. 
	Such behavior is a characteristic feature of non-diagonal bulk metrics and does not appear 
	in the case of diagonal metrics.\\
	Now using the expression of the tortoise coordinate in the above result, we get
	\begin{eqnarray}\label{eq14}
		h_{\omega}^{A}(r)&\approx&\frac{A_{1}\Bigg[1-i\omega\sqrt{g(\tilde{r}_{h})}R^2\int dr~\frac{1}{r^2 f(r)}\Bigg]}{\sqrt{\alpha(\tilde{r}_{h})}}~.
	\end{eqnarray}
	The expression can be further simplified by taking the near-horizon limit 
	$r \to \tilde r_h$. In this limit, we expand the blackening function $f(r)$ about 
	$r=\tilde r_h$ and retain only the leading term. Since we set $f(\tilde r_h)=0$, the 
	near-horizon behavior is governed by the linear contribution, yielding
	\begin{eqnarray}
		f(r)\approx(r-\tilde{r}_{h})f^{\prime}(\tilde{r}_{h})~.
	\end{eqnarray}
	Now substituting this form of $f(r)$ in eq.\eqref{eq14} and performing the integral we get 
	\begin{eqnarray}\label{EQ5}
		h_{\omega}^{A}(r)&\approx&\frac{A_1}{\sqrt{\alpha(\tilde{r}_{h})}}\left[1-\frac{i\omega R^2\sqrt{g(\tilde{r}_{h})}}{\tilde{r}_{h}^2f^{\prime}(\tilde{r}_{h})}\log\left(\frac{r}{\tilde{r}_{h}}-1\right)\right]~~~.
	\end{eqnarray}
	\subsubsection{Solution in the hydrodynamic limit ($\omega\rightarrow 0$)}
	\noindent Now we solve the equation of motion in the hydrodynamic limit, that is, in the limit $\omega\rightarrow 0$. In this limit we neglect terms which are proportional to $\omega^2$ and keep terms only up to order $\omega$. Keeping this in mind we recast eq.\eqref{Y} in the following form
	\begin{eqnarray}\label{EQ6}
		\partial_{r}\Bigg[\frac{\left(\frac{r}{R}\right)^4 f^{\frac{3}{2}}(r)}{\left(1-\gamma^2\left(\frac{r_{h}}{r}\right)^d\right)^{\frac{1}{2}}}~~\partial_{r}h_{\omega}^{B}\Bigg]=0~.
	\end{eqnarray}
	The most general solution of the above equation can be written as
	\begin{eqnarray}
		h_{\omega}^{B}(r)=B_{1}(\omega)\int dr\frac{\left(1-\gamma^2\left(\frac{r_{h}}{r}\right)^d\right)^{\frac{1}{2}}}{\left(\frac{r}{R}\right)^4 f^{\frac{3}{2}}(r)}+B_{2}(\omega)
	\end{eqnarray}
	where $B_{1}(\omega)$ and $B_{2}(\omega)$ are integration constants which depend only on $\omega$.
	Now we can approximate the above integral in the IR regime by using the fact $r\sim \tilde{r}_{h}$. In this regime the above solution can be written as 
	\begin{eqnarray}
		h_{\omega}^{B}(r)|_{\mathrm{IR}}\approx \frac{B_{1}(\omega)\left[1-\gamma^2\left(\frac{r_{h}}{\tilde{r}_{h}}\right)^d\right]^{\frac{1}{2}}}{\left(\frac{\tilde{r}_{h}}{R}\right)^4 f^{\prime}(\tilde{r}_{h})\sqrt{f(\tilde{r}_{h})}}\log\left(\frac{r}{\tilde{r}_h}-1\right)+B_{2}(\omega)~.
	\end{eqnarray}
	Now we can obtain $B_{1}(\omega)$ and $B_{2}(\omega)$ by comparing the above result with the solution given in eq.\eqref{EQ5}. This results
	\begin{eqnarray}\label{B2}
		B_{2}&=&\frac{A_{1}}{\sqrt{\alpha(\tilde{r}_{h})}}=\frac{A_{1}}{\left(\frac{\tilde{r}_{h}}{R}\right)}\frac{\left(1-\gamma^2\left(\frac{r_{h}}{\tilde{r}_{h}}\right)^d\right)^{\frac{1}{4}}}{f^{\frac{1}{4}}(\tilde{r}_{h})}~;~
		B_{1}=-i\omega A_{1}\frac{\tilde{r}_{h}}{R}\left(1+v^2\gamma^2\left(\frac{r_{h}}{\tilde{r}_{h}}\right)^d\right)\frac{f^{\frac{1}{4}}(\tilde{r}_{h})}{\left(1-\gamma^2\left(\frac{r_{h}}{\tilde{r}_{h}}\right)^d\right)^{\frac{1}{4}}}~.
	\end{eqnarray}
	One can determine the constant $A_{1}$ by normalization.
	\subsubsection{Solution in the UV limit}
	In the UV domain, that is, in the limit $r\rightarrow \infty$ the eq.\eqref{EQ1} reduces to 
	\begin{eqnarray}
		\partial_{r}\Bigg[\frac{\left(\frac{r}{R}\right)^4 f^{\frac{3}{2}}(r)}{\left(1-\gamma^2\left(\frac{r_{h}}{r}\right)^d\right)^{\frac{1}{2}}}~~\partial_{r}h_{\omega}^{C}\Bigg]=0~.
	\end{eqnarray}
	The above equation is very similar to that of the equation which descrbing the fluctuation of string the hydrodynamic limit as given in eq.\eqref{EQ6}. Therefore in this domain
	we utilize the solution obtained in the hydrodynamic limit. This yields
\begin{eqnarray}\label{uvs}
h_{\omega}^{C}(r)&=&B_{1}(\omega)\int dr\frac{\left(1-\gamma^2\left(\frac{r}{r_{h}}\right)^d\right)^{\frac{1}{2}}}{\left(\frac{r}{R}\right)^4 f^{\frac{3}{2}}(r)}+B_{2}(\omega)\nonumber\\
&\approx&\frac{A_{1}}{\left(\frac{\tilde{r}_{h}}{R}\right)}\frac{\left(1-\gamma^2\left(\frac{r_{h}}{\tilde{r}_{h}}\right)^d\right)^{\frac{1}{2}}}{f^{\frac{1}{4}}(\tilde{r}_{h})}\left[1-\frac{i\omega\left(\frac{\tilde{r}_{h}}{R}\right)^2f^{\frac{1}{2}}(\tilde{r}_{h})\left(1+v^2\gamma^2\left(\frac{r_{h}}{\tilde{r}_{h}}\right)^d\right)^{\frac{1}{2}}}{\left(1-\gamma^2\left(\frac{r_{h}}{\tilde{r}_{h}}\right)^d\right)^{\frac{1}{2}}}\int dr\frac{\left(1-\gamma^2\left(\frac{r}{r_{h}}\right)^d\right)^{\frac{1}{2}}}{\left(\frac{r}{R}\right)^4 f^{\frac{3}{2}}(r)}\right]~.\nonumber\\
\end{eqnarray}
In obtaining the above result, we have used the results given in eq.\eqref{B2}.
\subsection{Computation of admittance and diffusion coefficient}
\begin{figure}[!h]
				\begin{minipage}[t]{0.48\textwidth}
					\centering\includegraphics[width=\textwidth]{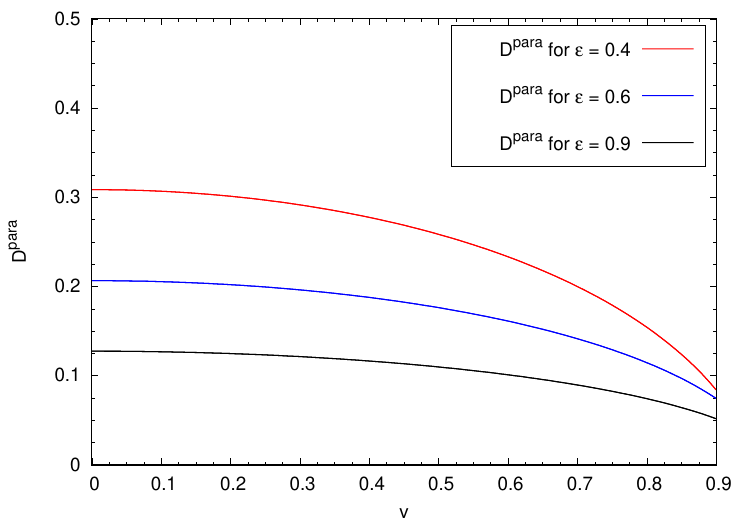}\\
					{\footnotesize Variation of $D^{para}$ w.r.t $v$ for different values of IR-cutoff} 
				\end{minipage}
				\begin{minipage}[t]{0.48\textwidth}
					\centering\includegraphics[width=\textwidth]{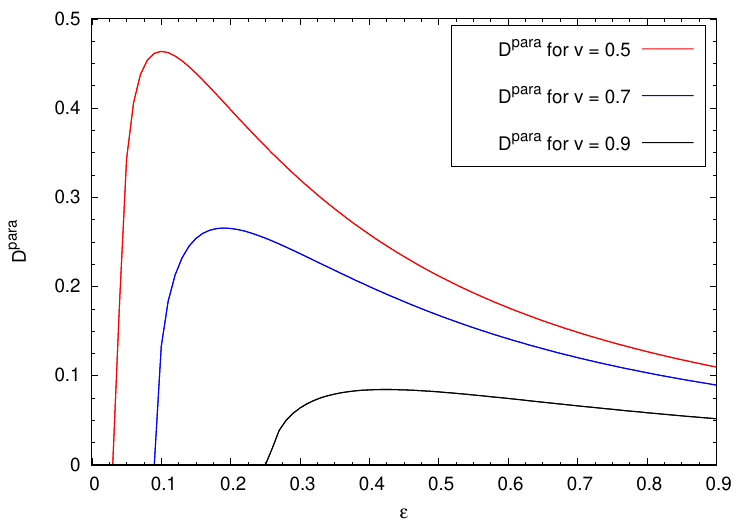}\\
					{\footnotesize Variation of $D^{para}$ w.r.t $\epsilon$ for different values of boost parameter.}
				\end{minipage}\hfill
				\caption{In the left pannel of the figure we have represented the variation of diffusion coefficient along the boost with the boost parameter. We have plotted the diffusion coeffcient for different values of the IR cutoff $\epsilon$. The red, blue, and black curves represent the variation of $D^{\text{para}}$ with the boost parameter for $\epsilon = 0.4$, $0.5$, and $0.9$, respectively. The right pannel of the above figure represents the variation of $D^{para}$ with the IR-cutoff for fixed value of boost parameter. The red, blue, and black curves represent the variation of $D^{\text{para}}$ with the IR cutoff for $v = 0.5$, $0.7$, and $0.9$, respectively.}
				
				\label{Diffpara}
			\end{figure}

With the above results in hand, we now proceed to compute the admittance and then the diffusion coefficient. One can define admittance as the linear response of a system against an external perturbation. Therefore, we need to apply an external force acting on the Brownian particle. Now from the holographic perspective one can introduce this external force by turning on an electromagnetic potential $A_{\mu}$ on the UV brane. This external electromagnetic field does not change the bulk dynamics, it only applies an external force to the string end point. The presence of the electromagnetic field modifies the Neumann boundary condition as we shall see in the discussion below. The modified Nambu-Goto action in the presence of an external electromagnetic field can be written as
\begin{eqnarray}
S_{NG}&\approx&-\frac{1}{2\pi \alpha^\prime}\int dt dr ~\sqrt{|g_{tt}|g_{rr}}~+~\frac{1}{4\pi \alpha ^\prime}\int dt dr \Bigg[\frac{\left(\partial_{r}Y(r,t)\right)^2}{2}\frac{(g_{tt}g_{yy}-g_{ty}^2)}{\sqrt{|g_{tt}|g{rr}}}+\frac{\left(\partial_{t}Y(r,t)\right)^2}{2}\frac{g_{yy}\sqrt{g_{rr}}}{\sqrt{|g_{tt}|}}\Bigg]\nonumber\\&+&\int dt~ (A_{t}
+\vec{A}.\vec{\dot{x}})|_{r=r_{b}}~.
\end{eqnarray}
Now varying the above action, we get the following modified equation of motion in the brane position \cite{PhysRevLett.110.061602,Giataganas:2018ekx,Caldeira:2020rir}
\begin{eqnarray}	F(t)&=&\frac{1}{2\pi\alpha^\prime}\Bigg(\frac{g_{tt}g_{yy}-g_{ty}^2}{\sqrt{|g_{tt}|g_{rr}}}\partial_{r}Y\Bigg)_{r=r_{b}}\nonumber\\
&=&\frac{1}{2\pi\alpha^\prime}\Bigg[\frac{\left(\frac{r}{R}\right)^4 f^{\frac{3}{2}}(r)}{\left(1-\gamma^2\left(\frac{r_{h}}{r}\right)^d\right)^{\frac{1}{2}}}~~\partial_{r}Y\Bigg]_{r=r_{b}}
\end{eqnarray}
where $F(t)$ is defined as $F(t)=\partial_{x}A_{t}-\partial_tA_{x}$.
To proceed further we have to substitute the metric component from eq.\eqref{geo} and the expression of $Y(r,t)=h_{\omega}^C(r) e^{i\omega t} $ ($h_{\omega}^C(r)$ is given by eq.\eqref{uvs}) in the above result. Using the form $F(t)=F(\omega)e^{i\omega t}$, we get
\begin{eqnarray}\label{f}
F(\omega)=\frac{1}{2\pi\alpha^\prime}\frac{-iA\omega\left(\frac{\tilde{r}_{h}}{R}\right)f^{\frac{1}{4}}(\tilde{r}_{h})\left(1+v^2\gamma^2\left(\frac{r_{h}}{\tilde{r}_{h}}\right)\right)^{\frac{1}{2}}}{\left(1-\gamma^2\left(\frac{r_{h}}{\tilde{r}_{h}}\right)^d\right)^{\frac{1}{4}}}~~~.
\end{eqnarray}
Keeping these results in mind, we can compute the admittance ($\xi(\omega)$) by using the following formula \cite{PhysRevLett.110.061602,Giataganas:2018ekx,Caldeira:2020rir}
\begin{equation}\label{adm}
\xi(\omega)=\frac{h_{\omega}^{C}(r_{b})}{F(\omega)}~.
\end{equation}
\noindent Now substituting the the expression of $h_{\omega}^C(r_{b})$ (given in eq.\eqref{uvs}) and $F(\omega)$ from eq.\eqref{f} in the above expression, we find the following form of the admittance
\begin{eqnarray}
\xi^{para}(\omega)=2\pi \alpha^{\prime}\frac{\frac{A_{1}}{\left(\frac{\tilde{r}_{h}}{R}\right)}\frac{\left(1-\gamma^2\left(\frac{r_{h}}{\tilde{r}_{h}}\right)^d\right)^{\frac{1}{2}}}{f^{\frac{1}{4}}(\tilde{r}_{h})}\left[1-\frac{i\omega\left(\frac{\tilde{r}_{h}}{R}\right)^2f^{\frac{1}{2}}(\tilde{r}_{h})\left(1+v^2\gamma^2\left(\frac{r_{h}}{\tilde{r}_{h}}\right)^d\right)^{\frac{1}{2}}}{\left(1-\gamma^2\left(\frac{r_{h}}{\tilde{r}_{h}}\right)^d\right)^{\frac{1}{2}}}\int dr\frac{\left(1-\gamma^2\left(\frac{r}{r_{h}}\right)^d\right)^{\frac{1}{2}}}{\left(\frac{r}{R}\right)^4 f^{\frac{3}{2}}(r)}\right]}{\frac{-iA\omega\left(\frac{\tilde{r}_{h}}{R}\right)f^{\frac{1}{4}}(\tilde{r}_{h})\left(1+v^2\gamma^2\left(\frac{r_{h}}{\tilde{r}_{h}}\right)\right)^{\frac{1}{2}}}{\left(1-\gamma^2\left(\frac{r_{h}}{\tilde{r}_{h}}\right)^d\right)^{\frac{1}{4}}}}~.
\end{eqnarray}
Therefore, the imaginary part of admittance reads
\begin{eqnarray}\label{imad}
\mathrm{Im}(\xi^{para}(\omega))&=&\frac{2\pi\alpha^\prime}{\left(\frac{\tilde{r}_{h}}{R}\right)^2\left(1+v^2\gamma^2\left(\frac{r_{h}}{\tilde{r}_{h}}\right)^d\right)^{\frac{1}{2}}}\frac{\left[1-\gamma^2\left(\frac{r_{h}}{\tilde{r}_{h}}\right)^d\right]^{\frac{1}{2}}}{f^{\frac{1}{2}}(\tilde{r}_{h})}.
\end{eqnarray}
Now the expression of the diffusion coefficient along the boost can be obtained by the following formula \cite{PhysRevLett.110.061602}
		\begin{equation}
			D^{para}=\frac{1}{\beta}~\mathop{\mathrm{lim}}_{\omega\rightarrow 0}\left(-i \omega\xi^{para}(\omega)\right)
		\end{equation}
		where $\beta$ is the inverse of the Hawking temperature ($\beta=\frac{1}{T}$) for the boosted AdS black hole, given by eq.\eqref{T}.\\ Thus, we can obtain the expression of the diffusion coefficient along the boost  by using the expression of admittance ($\xi^{para}(\omega)$) and inverse of Hawking temperature in the above result. This yields
\begin{equation}\label{d1}
D^{para}=\frac{2\pi\alpha^\prime T}{\left(\frac{\tilde{r}_{h}}{R}\right)^2\left(1+v^2\gamma^2\left(\frac{r_{h}}{\tilde{r}_{h}}\right)^d\right)^{\frac{1}{2}}}\frac{\left[1-\gamma^2\left(\frac{r_{h}}{\tilde{r}_{h}}\right)^d\right]^{\frac{1}{2}}}{f^{\frac{1}{2}}(\tilde{r}_{h})}~.
\end{equation} 
The above result suggests that we have obtained an expression of diffusion coefficient in terms of the black hole parameters. It is to be observed that in the absence of boost, that is,  in the limit $v\rightarrow 0$, one can get the general expression of diffusion coefficient for a generic diagonal metric as given in \cite{deBoer:2008gu}.
		The above result is a direct consequence of the classical version of the fluctuation dissipation theorem. To see this we have to rewrite the above result in the following form
		\begin{eqnarray}
			\frac{2\pi\alpha^\prime }{\left(\frac{\tilde{r}_{h}}{R}\right)^2\left(1+v^2\gamma^2\left(\frac{r_{h}}{\tilde{r}_{h}}\right)^d\right)^{\frac{1}{2}}}\frac{\left[1-\gamma^2\left(\frac{r_{h}}{\tilde{r}_{h}}\right)^d\right]^{\frac{1}{2}}}{f^{\frac{1}{2}}(\tilde{r}_{h})}=\frac{D^{para}}{T}~.\nonumber\\
		\end{eqnarray}
		The LHS of the above equation can also be interpreted as the mobility under the influcence of the external force on the Brownian particle.
		It is to be noted that in the limit $v\rightarrow 0$, we can recover the result of diffusion coefficient for AdS-Schwarzschild black hole in the bulk geometry. Notably, the diffusion coefficient decreases in the presence of a boost compared to the case without any boost which is depicted in the left panel of Figure\eqref{Diffpara}. Moreover, an increase in the boost leads to a further reduction in the diffusion coefficient, indicating a suppression of random motion with increasing drift. Another important observation from the left panel of Figure\eqref{Diffpara} is that, for a fixed boost parameter, the diffusion coefficient decreases as the cutoff is increased.
			\noindent Furthermore, it also to to be noted that in the right panel of Figure\eqref{Diffpara} we have represented the variation of diffusion coefficient as a function of the IR-cutoff for fixed valuse of the boost parameter. The plot suggests that starting from zero the diffusion coefficient increases with the IR-cutoff and recahes a maximum value then it agains starts to decreases. This observation holds for any fixed value of boost parameter.
\subsection{Correlation function and diffusion coefficient}
\noindent In this section, we will compute the diffusion coefficient by evaluating the thermal two-point correlation function of the string endpoints at the UV brane. To obtain the diffusion coefficient from the thermal two-point correlation function, we first need to calculate the mean square displacement, which quantifies the variance of the random walk of a particle. In the following, we will calculate this mean square displacement in a general setup, utilizing the results from the previous sections.
To obtain the mean square displacement, we need to impose the ingoing and outgoing boundary condition near the horizon.\\
The general solution for the string fluctuation in the IR region can be written as
\begin{eqnarray}\label{genir}
h_{\omega}^{\mathrm{IR}}(r)=A\left[e^{i \frac{\omega}{\gamma}\sqrt{g(\tilde{r}_{h})} r_{*}}+B~e^{-i \frac{\omega}{\gamma}\sqrt{g(\tilde{r}_{h})} r_{*}}\right]~.
\end{eqnarray}
On the other hand, in the UV region the solution of string fluctuation for small frequency can be written as 
\begin{eqnarray}\label{genuv}
h_{\omega}^{\mathrm{UV}}(r)&=&A\Bigg[\left(1-\frac{i\omega\left(\frac{\tilde{r}_{h}}{R}\right)^2f^{\frac{1}{2}}(\tilde{r}_{h})\left(1+v^2\gamma^2\left(\frac{r_{h}}{\tilde{r}_{h}}\right)^d\right)^{\frac{1}{2}}}{\left(1-\gamma^2\left(\frac{r_{h}}{\tilde{r}_{h}}\right)^d\right)^{\frac{1}{2}}}\int dr\frac{\left(1-\gamma^2\left(\frac{r}{r_{h}}\right)^d\right)^{\frac{1}{2}}}{\left(\frac{r}{R}\right)^4 f^{\frac{3}{2}}(r)}\right)\nonumber\\&+&B\left(1+\frac{i\omega\left(\frac{\tilde{r}_{h}}{R}\right)^2f^{\frac{1}{2}}(\tilde{r}_{h})\left(1+v^2\gamma^2\left(\frac{r_{h}}{\tilde{r}_{h}}\right)^d\right)^{\frac{1}{2}}}{\left(1-\gamma^2\left(\frac{r_{h}}{\tilde{r}_{h}}\right)^d\right)^{\frac{1}{2}}}\int dr\frac{\left(1-\gamma^2\left(\frac{r}{r_{h}}\right)^d\right)^{\frac{1}{2}}}{\left(\frac{r}{R}\right)^4 f^{\frac{3}{2}}(r)}\right)\Bigg]~~~.
\end{eqnarray}
We have assumed that both the constants $A$ and $B$ are same in these two different regions. One of these constants (in particular $B$) can be obtained by applying the Neumann boundary condition (given by eq.\eqref{nbc}) and  the other constant $A$ can be computed by normalization.\subsubsection{Neumann boundary condition and normalization}\label{AB}
\noindent As already mentioned, one can obtain the explicit form of the constant 
$B$ by applying the Neumann boundary condition near the horizon. On the other hand, to determine the explicit form of  $A$, we must use the procedure of normalization.\\
First, we compute the explicit form of $B$ by applying the Neumann boundary condition at the UV cutoff, which is characterized by $r=r_{b}$. It can be shown that at leading order in $\omega$, this constant simplifies to 
\begin{equation}
B=1~.
\end{equation}
On the other hand, in general for higher order in $\omega$, this coefficient $B$
is a pure phase factor, $e^{i\omega \theta}$, for some real $\theta$. To get this result let us rewrite the  UV solution given in eq.\eqref{genuv} in the following form 
\begin{eqnarray}
h_{\omega}^{\mathrm{UV}}(r)=A[G(r)+B~G^{*}(r)]
\end{eqnarray}
where $g^{*}(r)$ is the complex conjugate of $g(r)$. Now applying the Neumann boundary condition at the UV cutoff $r=r_b$, we  get
\begin{equation}
B=-\left(\frac{\partial_{r} G(r)}{\partial_{r}G^{*}(r)}\right)_{r=r_{b}}=e^{i\omega \theta}~.
\end{equation} 
On the other hand, applying the Neumann boundary condition near the horizon we get the following form of $B$
\begin{eqnarray}\label{Bir}
B=e^{2i \frac{\omega}{\gamma}\sqrt{g(\tilde{r}_{h})}}|_{r=\tilde{r}_{h}(1+\epsilon)}
\end{eqnarray}
where $\epsilon$ is the IR cutoff, $\epsilon<<1$. In order to get the above result we have used the IR solution given in eq.\eqref{genir}. Before proceeding further, it is important to note that we impose the Neumann boundary condition in the near-horizon region. If one instead imposes the Dirichlet boundary condition at the horizon, the solution becomes trivial, leading to the absence of non-trivial modes in the IR region. Therefore, in the present analysis, the Neumann boundary condition must be imposed in the near-horizon sector. This choice has also been discussed in \cite{deBoer:2008gu}.
Now, in the near horizon limit the expression of the tortoise coordinate reads
\begin{equation}
r_{*}\approx\frac{\gamma R^2}{\tilde{r}_{h}^2f^{\prime}(\tilde{r}_{h})}\log\left(\frac{r}{\tilde{r}_{h}}-1\right)~.
\end{equation}
Now using the above result in eq.\eqref{Bir}, we get
\begin{eqnarray}
B=\exp(-\frac{2i\omega R^2\sqrt{g(\tilde{r}_{h})}}{f^{\prime}(\tilde{r}_{h}) \tilde{r}_{h}^2}\log(\frac{1}{\epsilon}))~.
\end{eqnarray}
The above result suggests that $B$ is a pure phase in $\omega$. As $\epsilon<<1$, this condition implies that the frequency spectrum is no longer continuous but instead becomes discrete. This discreteness of the frequencies are given by
\begin{eqnarray}
\Delta\omega=\frac{\pi \left(\frac{\tilde{r}_{h}}{R}\right)^2f^{\prime}(\tilde{r}_{h})}{\log(\frac{1}{\epsilon})\left(1+v^2\gamma^2\left(\frac{r_{h}}{\tilde{r}_{h}}\right)^d\right)^{\frac{1}{2}}}~.
\end{eqnarray}
Hence, the density of states can be written as 
\begin{eqnarray}
\mathcal{D}(\omega)=\frac{1}{\Delta \omega}=\frac{\left(1+v^2\gamma^2\left(\frac{r_{h}}{\tilde{r}_{h}}\right)^d\right)^{\frac{1}{2}}\log(\frac{1}{\epsilon})}{\pi \left(\frac{\tilde{r}_{h}}{R}\right)^2f^{\prime}(\tilde{r}_{h})}~.
\end{eqnarray}
The preceding result plays a crucial role in the subsequent derivation of the thermal two-point correlation function.\\
Now we have to obtain an explicit form of $A$ by using the normalization method.  To do this let us first define the Klein-Gordon inner product as follows
\begin{eqnarray}\label{kgp}
(f,g)=-\frac{i}{2\pi\alpha^\prime}\int_{\Sigma} dx \sqrt{h} n^\mu g_{yy}(f\partial_{\mu} g^{*}-g^{*}\partial_{\mu}f)
\end{eqnarray}
where $h$ is the determinant of the induced metric on the Cauchy slice $\Sigma$, $n^\mu$ is the unit normal on this surface $\Sigma$, and $f$ and $g$ are solutions of the equation of motion. In this scenario, we can choose the Cauchy surface to be a $t$ and $r$ part of the bulk metric given in eq.\eqref{geo}. Therefore, on this surface the expression of the unit normal and determinant of the induced metric is given by
\begin{eqnarray}
n^\mu=\left(\frac{1}{\sqrt{|g_{tt}|}},0\right)~~;~~h=g_{rr}~.
\end{eqnarray}
Now substituting the above result in eq.\eqref{kgp}, the Klein-Gordon inner product can be written as 
\begin{eqnarray}
(f,g)=-\frac{i}{2\pi \alpha^\prime}\int dr \sqrt{\frac{g_{rr}}{g_{tt}}} g_{yy}(f\partial_{\mu} g^{*}-g^{*}\partial_{\mu}f)~.
\end{eqnarray}
To proceed further, we define $f$ and $g$ as follows
\begin{eqnarray}\label{fg}
f(r,t)=Y(r,t)=h_{\omega}(r) e^{i\omega t}\nonumber\\
g(r,t)=Y(r,t)=h_{\omega}(r)e^{i\omega t}~.
\end{eqnarray}
Now a proper normalization can be defined as
\begin{eqnarray}
\left(Y(r,t),Y(r,t)\right)=1
\end{eqnarray}
Substituting the form of $Y(r,t)$ given by eq.\eqref{fg} in the the above result, we get
\begin{eqnarray}
\frac{\omega}{\pi\alpha^\prime}\int dr \sqrt{\frac{g_{rr}}{g_{tt}}} g_{yy}(2i\omega |h_{\omega}(r)|^2)=1~.
\end{eqnarray}
\noindent Now  to get an explicit form of $A$ we have to substitute the expression of $h_{\omega}(r)$ (given by eq.\eqref{genir}) in the above result. This yields 
\begin{eqnarray}
|A|^2&\approx&\frac{\pi \alpha^\prime f^{\prime}(\tilde{r}_{h})\left(1-\gamma^2\left(\frac{r_{h}}{\tilde{r}_{h}}\right)^d\right)^{\frac{1}{2}}}{\omega\left(1+v^2\gamma^2\left(\frac{r_{h}}{\tilde{r}_{h}}\right)^d\right)\sqrt{f(\tilde{r}_{h})}\log(\frac{1}{\epsilon})}
=\frac{\alpha^\prime\left(1-\gamma^2\left(\frac{r_{h}}{\tilde{r}_{h}}\right)^d\right)^{\frac{1}{2}}}{\left(\frac{\tilde{r}_{h}}{R}\right)^2\left(1+v^2\gamma^2\left(\frac{r_{h}}{\tilde{r}_{h}}\right)^d\right)^{\frac{1}{2}}\sqrt{f(\tilde{r}_{h})}}\frac{\Delta\omega}{\omega}~.
\end{eqnarray}
It is important to note that in obtaining the above result, terms up to linear order in $\omega$ and quadratic order in the boost $v$ have been considered. \\

\subsubsection{Correlation function and mean square displacement}
\noindent With the above results in hand, we now proceed to compute the thermal correlation function of the string endpoints. To this end, we first quantize the string fluctuations. This is achieved by expressing the solution of eq.\eqref{stf} in terms of creation and annihilation operators in the vicinity of the boundary. This leads to the following result
\begin{eqnarray}
Y(r,t)=\sum_{\omega>0}\left[a_{\omega}h_{\omega}^{\mathrm{UV}}(r)e^{-i\omega t}+a_{\omega}^{\dagger}(h_{\omega}^{\mathrm{UV}}(r))^{*}e^{i\omega t}\right]
\end{eqnarray}
where $a_{\omega}$ and $a_{\omega}^{\dagger}$ are annihilation and creation operators with the following commutation relations
\begin{eqnarray}
[a_{\omega},a_{\omega^{\prime}}^{\dagger}]=\delta_{\omega\omega^{\prime}}~~;~~	[a_{\omega}^{\dagger},a_{\omega^{\prime}}^{\dagger}]=	[a_{\omega},a_{\omega^{\prime}}]=0~~~.
\end{eqnarray}
We compute this correlation functions by considering the canonical ensemble. The density matrix for the canonical ensemble reads
\begin{eqnarray}\label{den}
\rho_{0}=\frac{e^{-\beta\sum_{\omega>0}\omega a_{\omega}^{\dagger}a_{\omega}}}{Tr(e^{-\beta\sum_{\omega>0}\omega a_{\omega}^{\dagger}a_{\omega}})}
\end{eqnarray}
where  $a_{\omega}$ and $a_{\omega}^{\dagger}$ are annihilation and creation operator respectively which obey
\begin{eqnarray}\label{expv}
\langle a_{\omega}^{\dagger}a_{\omega^{\prime}}\rangle_{\rho_{0}}&=&\frac{\delta_{\omega\omega^{\prime}}}{e^{\beta\omega}\pm 1}~;~\langle a_{\omega}^{\dagger}a_{\omega^{\prime}}^{\dagger}\rangle_{\rho_{0}}=	\langle a_{\omega}a_{\omega^{\prime}}\rangle_{\rho_{0}}=0~
\end{eqnarray}
where plus (minus) sign represents the fermionic (bosonic) statistics. Before proceeding further, we would like to make a few comments regarding the bosonic and fermionic cases. Here, the string endpoint, which represents the boundary particle, is still bosonic; however, the plasma in which the particle is immersed may behave as either fermionic or bosonic manner. Therefore, in the following sections, we consider these two cases separately.\\
\noindent Now let us define the mean square displacement $s^{2}(t)$ as
\begin{eqnarray}\label{s}
s^{2}(t)&\equiv&\langle[y(t)-y(0)]^2\rangle\nonumber\\
&=&\langle[Y(t,r_{b})-Y(0,r_{b})]^2\rangle~~
\end{eqnarray}
where $y(t)=Y(t,r_{b})$ represents the position of the boundary ($r=r_{b}$) particle at time $t$.
Now let us compute the correlation between position of the boundary particle at two different time $t_{1}$ and $t_{2}$, that is, $\langle y(t_{1})y(t_{2})\rangle$
\begin{eqnarray}
\langle y(t_{1})y(t_{2})\rangle&=&\langle Y(t_{1},r_{b})Y(t_{2},r_{b})\rangle\nonumber\\
&=&\sum_{\omega,\omega^\prime>0}\Bigg(\langle a_{\omega}^{\dagger}a_{\omega^{\prime}}\rangle_{\rho_{0}}(h_{\omega}^{\mathrm{UV}*}(r_{b})h_{\omega^{\prime}}^{\mathrm{UV}}(r_{b}))e^{i\omega t_{1}-i\omega^{\prime}t_{2}}\nonumber+\langle a_{\omega}a_{\omega^{\prime}}^{\dagger}\rangle_{\rho_{0}}(h_{\omega}^{\mathrm{UV}}(r_{b})h_{\omega^{\prime}}^{\mathrm{UV}*}(r_{b}))e^{-i\omega t_{1}+i\omega^{\prime}t_{2}}\Bigg)\nonumber\\	&=&\sum_{\omega>0}\left[|h_{\omega}(r_{b})|^2\frac{2\cos(\omega(t_{1}-t_{2}))}{e^{\beta\omega}\pm1}+|h_{\omega}(r_{b})|^2 e^{i\omega(t_{2}-t_{1})}\right]~.
\end{eqnarray}
\noindent To get the last line we have used the results given in eq.\eqref{expv}. Further, substituting  $t_{1}=t$ and $t_{2}=0$ in the above result, we get
	\begin{eqnarray}\label{corr1}
	\langle y(t)y(0)\rangle&=&\sum_{\omega>0}|h_{\omega}^{\mathrm{UV}}(r_{b})|^2\left(\frac{2 \cos(\omega t)}{e^{\beta\omega}\pm1}+e^{-i\omega t}\right)\nonumber\\
	&=&\frac{\alpha^\prime\left(1-\gamma^2\left(\frac{r_{h}}{\tilde{r}_{h}}\right)^d\right)^{\frac{1}{2}}}{\left(\frac{\tilde{r}_{h}}{R}\right)^2\left(1+v^2\gamma^2\left(\frac{r_{h}}{\tilde{r}_{h}}\right)^d\right)^{\frac{1}{2}}\sqrt{f(\tilde{r}_{h})}}\sum_{\omega>0}\frac{\Delta\omega}{\omega}\left(\frac{2 \cos(\omega t)}{e^{\beta\omega}\pm1}+e^{-i\omega t}\right)~.
	\end{eqnarray}
In the second line of the above equation we have substituted the expression for $h_{\omega}^{\mathrm{UV}}(r_{b})$ from eq.\eqref{genuv} and kept terms only up to $\mathcal{O}\left(\frac{1}{\omega^2}\right)$. This implies that as we are in the small frequency domain.
\noindent Furthermore, we can approximate the sum by definite integral by using the following relation
\begin{eqnarray}
\sum_{\omega>0}	\Delta\omega\rightarrow\int_{0}^{\infty}d\omega\Leftrightarrow\sum_{\omega>0}\frac{\pi \left(\frac{\tilde{r}_{h}}{R}\right)^2f^{\prime}(\tilde{r}_{h})}{\log(\frac{1}{\epsilon})\left(1+v^2\gamma^2\left(\frac{r_{h}}{\tilde{r}_{h}}\right)^d\right)^{\frac{1}{2}}}\rightarrow\int_{0}^{\infty} d\omega~.\nonumber\\
\end{eqnarray}
Therefore, the integral form of eq.\eqref{corr1} reads
\begin{eqnarray}\label{corr2}
\langle y(t)y(0)\rangle&=&\frac{\alpha^\prime\left(1-\gamma^2\left(\frac{r_{h}}{\tilde{r}_{h}}\right)^d\right)^{\frac{1}{2}}}{\left(\frac{\tilde{r}_{h}}{R}\right)^2\left(1+v^2\gamma^2\left(\frac{r_{h}}{\tilde{r}_{h}}\right)^d\right)^{\frac{1}{2}}\sqrt{f(\tilde{r}_{h})}}\int_{0}^{\infty}\frac{d \omega}{\omega}\left(\frac{2 \cos(\omega t)}{e^{\beta\omega}\pm1}+e^{-i\omega t}\right)\nonumber\\
&=&\langle y(0)y(t)\rangle^*~.
\end{eqnarray}
On the other hand, the integral form of the auto-correlation function reads
\begin{eqnarray}\label{corr3}
\langle y(t)y(t)\rangle&=&\frac{\alpha^\prime\left(1-\gamma^2\left(\frac{r_{h}}{\tilde{r}_{h}}\right)^d\right)^{\frac{1}{2}}}{\left(\frac{\tilde{r}_{h}}{R}\right)^2\left(1+v^2\gamma^2\left(\frac{r_{h}}{\tilde{r}_{h}}\right)^d\right)^{\frac{1}{2}}\sqrt{f(\tilde{r}_{h})}}\int_{0}^{\infty}\frac{d \omega}{\omega}\left(\frac{2}{e^{\beta\omega}\pm1}+1\right)\nonumber\\&=&\langle y(0)y(0)\rangle~.
\end{eqnarray}
Substituting the above results (given in eq.\eqref{corr2} and eq.\eqref{corr3})  in eq.\eqref{s}, one can compute the explicit form of the mean square displacement. But this expression of mean square displacement has a divergence. This motivates us to define a regularized version of mean square displacement. Therefore, one can define the regularized mean square displacement (RMSD) as 
\begin{eqnarray}\label{reg}
s^{2}_{reg}(t)|^{para}&\equiv&\langle:[y(t)-y(0)]^2:\rangle=\langle:[Y(t,r_{b})-Y(0,r_{b})]^2:\rangle~~~~~
\end{eqnarray}
In the above equation $:[...]:$ represents the normal ordering. We can further simplify the above expression of RMSD. This results in
\begin{eqnarray}
s^{2}_{reg}(t)|^{para}=2\langle:y^{2}(t):\rangle-2\langle:y(t)y(0):\rangle~.
\end{eqnarray}
The above expression suggests that we need to compute $\langle :y^{2}(t):\rangle $ and $\langle:y(t)y(0):\rangle$ to get an explicit form of $s^{2}_{reg}(t)$.
\noindent The expression of  $\langle:y(t)y(0):\rangle$ and $\langle :y^{2}(t):\rangle $  can be obtained  from eq.\eqref{corr2} and eq.\eqref{corr3} by considering normal ordering. This results in 
\begin{eqnarray}
\langle:y^{2}(t):\rangle&=&\frac{\alpha^\prime\left(1-\gamma^2\left(\frac{r_{h}}{\tilde{r}_{h}}\right)^d\right)^{\frac{1}{2}}}{\left(\frac{\tilde{r}_{h}}{R}\right)^2\left(1+v^2\gamma^2\left(\frac{r_{h}}{\tilde{r}_{h}}\right)^d\right)^{\frac{1}{2}}\sqrt{f(\tilde{r}_{h})}}\int_{0}^{\infty}\frac{d\omega}{\omega}\frac{2}{e^{\beta\omega}\pm1}=\langle:y^{2}(0):\rangle~~~\\
\langle:y(t)y(0):\rangle&=&\frac{\alpha^\prime\left(1-\gamma^2\left(\frac{r_{h}}{\tilde{r}_{h}}\right)^d\right)^{\frac{1}{2}}}{\left(\frac{\tilde{r}_{h}}{R}\right)^2\left(1+v^2\gamma^2\left(\frac{r_{h}}{\tilde{r}_{h}}\right)^d\right)^{\frac{1}{2}}\sqrt{f(\tilde{r}_{h})}}\int_{0}^{\infty}\frac{d\omega}{\omega}\frac{2 \cos(\omega t)}{e^{\beta\omega}\pm1}~~.
\end{eqnarray}
Substituting the above results in eq.\eqref{reg}, we can obtain divergent free expression of mean square displacement. This yields
	\begin{eqnarray}\label{sgen}
	s^{2}_{reg}(t)|^{para}=\frac{8\alpha^\prime\left(1-\gamma^2\left(\frac{r_{h}}{\tilde{r}_{h}}\right)^d\right)^{\frac{1}{2}}}{\left(\frac{\tilde{r}_{h}}{R}\right)^2\left(1+v^2\gamma^2\left(\frac{r_{h}}{\tilde{r}_{h}}\right)^d\right)^{\frac{1}{2}}\sqrt{f(\tilde{r}_{h})}}\int_{0}^{\infty}\frac{\sin[2](\frac{\omega t}{2})}{e^{\beta\omega}\pm1}~.
	\end{eqnarray}
Keeping this general expression of RMSD in mind, we now proceed to compute $s^{2}_{reg}(t)$ for various interesting scenarios in the following sections.
\subsubsection{Computation of regularized mean square displacement in different scenarios}
\begin{figure}[!h]
\centering
\includegraphics[width=0.5\textwidth]{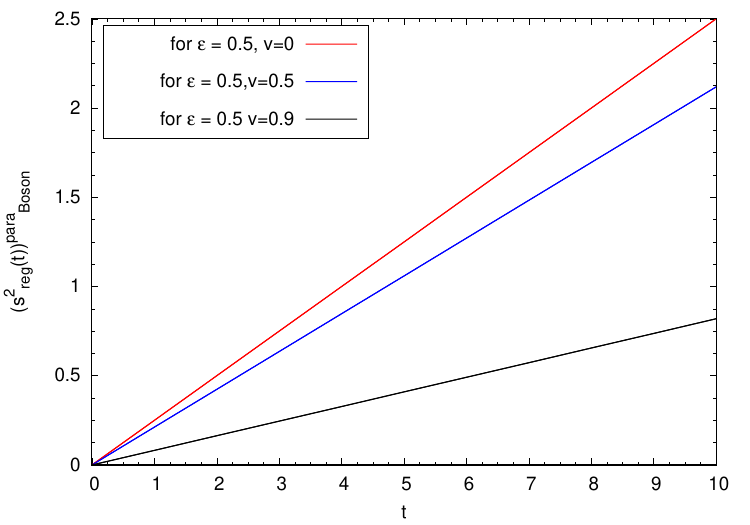}
\caption{The Figure above shows the variation of $(s^{2}_{\mathrm{reg}}(t))^{\mathrm{para}}_{\mathrm{Boson}}$ with respect to the observer’s time $t$. For this plot, we have considered the parameters $d = 4$, $\alpha = 10$,$\epsilon=0.5$ and $r_h = 20$. The red, blue, and black curves represent the RMSD results (as given in Eq.~\eqref{diffpara}) for boost velocities $v = 0$, $0.5$, and $0.9$, respectively.}
\label{rmsdpara}
\end{figure}

\noindent This section presents the computation of the regularized mean square displacement (RMSD) in various scenarios, including both bosonic and fermionic cases. 
For the bosonic case, the RMSD can be derived by adopting the negative sign in the denominator of the previously obtained expression. This yields
\begin{eqnarray}
(s^{2}_{reg}(t))^{para}_{\mathrm{Boson}}&=&\frac{8\alpha^\prime\left(1-\gamma^2\left(\frac{r_{h}}{\tilde{r}_{h}}\right)^d\right)^{\frac{1}{2}}}{\left(\frac{\tilde{r}_{h}}{R}\right)^2\left(1+v^2\gamma^2\left(\frac{r_{h}}{\tilde{r}_{h}}\right)^d\right)^{\frac{1}{2}}\sqrt{f(\tilde{r}_{h})}}\int_{0}^{\infty}\frac{\sin[2](\frac{\omega t}{2})}{e^{\beta\omega}-1}\nonumber\\
&=&\frac{2\alpha^\prime\left(1-\gamma^2\left(\frac{r_{h}}{\tilde{r}_{h}}\right)^d\right)^{\frac{1}{2}}}{\left(\frac{\tilde{r}_{h}}{R}\right)^2\left(1+v^2\gamma^2\left(\frac{r_{h}}{\tilde{r}_{h}}\right)^d\right)^{\frac{1}{2}}\sqrt{f(\tilde{r}_{h})}}\log\left(\frac{\sinh(\frac{t\pi}{\beta})}{\left(\frac{t\pi}{\beta}\right)}\right)~.
\end{eqnarray}
Keeping this result in mind, we now proceed to find the expression of RMSD for both the early and late time domains. In the early time domain (also referred to ballistic time domain), that is, for $t<<\beta$,  the expression of RMSD reads
\begin{eqnarray}
(s^{2}_{reg}(t))^{para}_{\mathrm{Boson}}\approx\frac{2\alpha^\prime\left(1-\gamma^2\left(\frac{r_{h}}{\tilde{r}_{h}}\right)^d\right)^{\frac{1}{2}}}{\left(\frac{\tilde{r}_{h}}{R}\right)^2\left(1+v^2\gamma^2\left(\frac{r_{h}}{\tilde{r}_{h}}\right)^d\right)^{\frac{1}{2}}\sqrt{f(\tilde{r}_{h})}}\left(\frac{\pi}{\beta}\right)^2 t^2~~~~~.
\end{eqnarray}
The above result suggests that in the ballistic time domain, the RMSD increase quadratically with time.
On the other hand, in the late time domain (also known as the diffusive regime), that is, for $t>>\beta$, we have
\begin{eqnarray}\label{diffpara}
(s^{2}_{reg}(t))^{para}_{\mathrm{Boson}}&\approx& \frac{2\pi\alpha^\prime\left(1-\gamma^2\left(\frac{r_{h}}{\tilde{r}_{h}}\right)^d\right)^{\frac{1}{2}}}{\left(\frac{\tilde{r}_{h}}{R}\right)^2\left(1+v^2\gamma^2\left(\frac{r_{h}}{\tilde{r}_{h}}\right)^d\right)^{\frac{1}{2}}\sqrt{f(\tilde{r}_{h})}} \left(\frac{t}{\beta}\right)
=D^{para} t~.
\end{eqnarray}
In the above result $D^{para}$ is nothing but the diffusion coefficient along the direction of boost given by
\begin{eqnarray}
D^{para}=\frac{2\alpha^\prime\left(1-\gamma^2\left(\frac{r_{h}}{\tilde{r}_{h}}\right)^d\right)^{\frac{1}{2}}T}{\left(\frac{\tilde{r}_{h}}{R}\right)^2\left(1+v^2\gamma^2\left(\frac{r_{h}}{\tilde{r}_{h}}\right)^d\right)^{\frac{1}{2}}\sqrt{f(\tilde{r}_{h})}}
\end{eqnarray}
It is also to be noted that the result of diffusion coefficient along the direction of boost computed above, matches exactly with our earlier result given in eq.\eqref{d1}. We have represented the this result of RMSD for late time graphically for different values of boost in Fig.\eqref{rmsdpara}. This Figure suggests that the RMSD decreases with increasing boost.
On the other hand, the slope of the $(s^{2}{\mathrm{reg}}(t))^{\mathrm{para}}_{\mathrm{Boson}}$ vs. $t$ graph represents the diffusion coefficient. This graph also indicates that the slope decreases as the boost increases, implying that the diffusion process slows down with higher boost.
Now for the fermioinc plasma, the expression of $s^{2}_{reg}(t)$ reads
\begin{eqnarray}
(s^{2}_{reg}(t))^{para}_{\mathrm{Fermion}}&=&\frac{8\alpha^\prime\left(1-\gamma^2\left(\frac{r_{h}}{\tilde{r}_{h}}\right)^d\right)^{\frac{1}{2}}}{\left(\frac{\tilde{r}_{h}}{R}\right)^2\left(1+v^2\gamma^2\left(\frac{r_{h}}{\tilde{r}_{h}}\right)^d\right)^{\frac{1}{2}}\sqrt{f(\tilde{r}_{h})}}\int_{0}^{\infty}\frac{\sin[2](\frac{\omega t}{2})}{e^{\beta\omega}+1}\nonumber\\
&=&\frac{2\alpha^\prime\left(1-\gamma^2\left(\frac{r_{h}}{\tilde{r}_{h}}\right)^d\right)^{\frac{1}{2}}}{\left(\frac{\tilde{r}_{h}}{R}\right)^2\left(1+v^2\gamma^2\left(\frac{r_{h}}{\tilde{r}_{h}}\right)^d\right)^{\frac{1}{2}}\sqrt{f(\tilde{r}_{h})}}\log(\frac{\frac{t\pi}{2\beta}}{\tanh(\frac{t\pi}{2\beta})})~.
\end{eqnarray}
Now in the early time domain, that is, the domain $t<<\beta$, the above expression reduces to
\begin{eqnarray}\label{fpar}
(s^{2}_{reg}(t))^{para}_{\mathrm{Fermion}}&\sim&\frac{\pi^2\alpha^\prime\left(1-\gamma^2\left(\frac{r_{h}}{\tilde{r}_{h}}\right)^d\right)^{\frac{1}{2}}}{\left(\frac{\tilde{r}_{h}}{R}\right)^2\left(1+v^2\gamma^2\left(\frac{r_{h}}{\tilde{r}_{h}}\right)^d\right)^{\frac{1}{2}}\sqrt{f(\tilde{r}_{h})}}\left(\frac{t}{\beta}\right)^2~.
\end{eqnarray}
This above result again agrees with that of the ballistic regime. On the other hand in the late time domain, that is, for $t>>\beta$, we have
\begin{eqnarray}
(s^{2}_{reg}(t))^{para}_{\mathrm{Fermion}}&\sim&\frac{\alpha^\prime\left(1-\gamma^2\left(\frac{r_{h}}{\tilde{r}_{h}}\right)^d\right)^{\frac{1}{2}}}{\left(\frac{\tilde{r}_{h}}{R}\right)^2\left(1+v^2\gamma^2\left(\frac{r_{h}}{\tilde{r}_{h}}\right)^d\right)^{\frac{1}{2}}\sqrt{f(\tilde{r}_{h})}}\log(\frac{\pi t}{2\beta})~.
\end{eqnarray}
\begin{figure}[!h]
\centering
\includegraphics[width=0.5\textwidth]{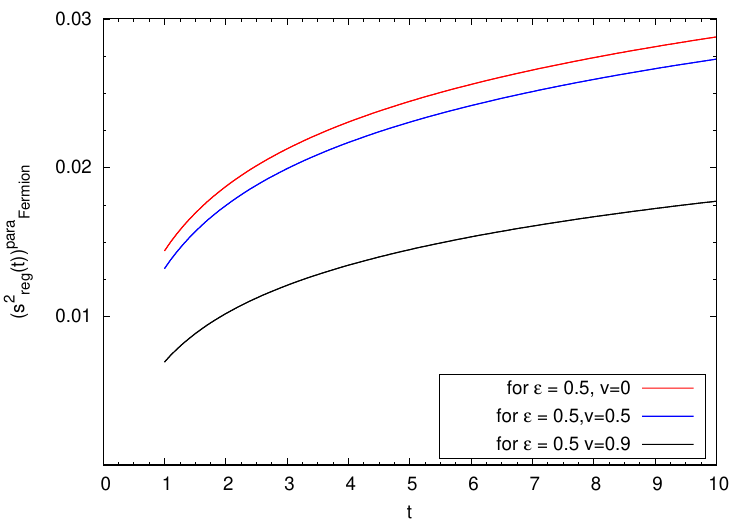}
\caption{The Figure above shows the variation of $(s^{2}{\mathrm{reg}}(t))^{\mathrm{para}}_{\mathrm{Fermion}}$ with respect to the observer’s time $t$. For this plot, we have considered the parameters $d = 4$, $\alpha = 10$, and $r_h = 20$. The red, blue, and black curves represent the RMSD results (as given in Eq.~\eqref{fpar}) for boost velocities $v = 0$, $0.5$, and $0.9$, respectively.}
\label{RMSDparafermion}
\end{figure}
	Now we would like to make few comments on the above results. We have observed that, in the ballistic time domain, $s^{2}_{reg}(t)|^{para}$ increases quadratically with time for both the bosonic and fermioinc case. However, in the late time time domain $s^{2}_{reg}(t)|^{para}$ behaves differently for bosons and fermions. In particular for the bosonic case, RMSD varies linearly with time, which represents usual diffusion. On the other hand for fermions, we have found that, RMSD varies as $\sim \log(t)$.  Therefore, for the fermions the diffusion process occurs very slowly. This phenomena is referred to the Sinai-like diffusion which is represented in Fig.\eqref{RMSDparafermion}.
\subsection{Fluctuation-dissipation theorem}\label{FDpara}
\noindent The fluctuation-dissipation theorem (FDT) is a key result in statistical mechanics that connects the fluctuations of a system in equilibrium to its response to external perturbations. It establishes a relationship between equilibrium thermodynamic properties (such as fluctuations) and transport properties (such as dissipation). The fluctuation-dissipation theorem can be applied to various systems, including both bosons and fermions.
In this discussion, we will explore the fluctuation-dissipation theorem in the context of both bosonic and fermionic systems. This analysis also offers a systematic approach to verify whether the results obtained so far are consistent or not. By applying the fluctuation-dissipation theorem (FDT) to both bosonic and fermionic systems with a chemical potential, we can cross-check the relationships between equilibrium fluctuations and response functions. \\
\noindent To proceed further let us construct a symmetric Green's function of the following form
\begin{eqnarray}
G_{\mathrm{sym}}^{\mathrm{B,F}}=\frac{1}{2}\left(\langle y(t)y(0)\rangle+\langle y(0)y(t)\rangle\right)
\end{eqnarray}
where $\mathrm{B ~and ~F}$ stands for bosons and fermions respectively. One can compute this symmetric Green's function by substituting the expression of $\langle y(t)y(0)\rangle $(given in eq.\eqref{corr2}) in the above result. This yields	
\begin{eqnarray}	G_{\mathrm{sym}}^{\mathrm{B,F}}=\frac{8\alpha^\prime\left(1-\gamma^2\left(\frac{r_{h}}{\tilde{r}_{h}}\right)^d\right)^{\frac{1}{2}}}{\left(\frac{\tilde{r}_{h}}{R}\right)^2\left(1+v^2\gamma^2\left(\frac{r_{h}}{\tilde{r}_{h}}\right)^d\right)^{\frac{1}{2}}\sqrt{f(\tilde{r}_{h})}}~\int_{-\infty}^{\infty}\frac{d\omega}{|\omega|}\left(\frac{2}{e^{\beta|\omega|}\pm 1}+1\right)e^{i\omega t}~.
\end{eqnarray}
In earlier studies it was shown that the FDT for bosons and ferimons can be written as \cite{Markov:2009ue}
\begin{eqnarray}
G_{\mathrm{sym}}^{\mathrm{B,F}}=\mathcal{F}^{-1}\left[(1+2 n_{n_{\mathrm{B,F}}})\mathrm{Im}(\xi(\omega))\right]
\end{eqnarray}
where $n_{\mathrm{B,F}}$ represents the Bose-Einstein distribution (for bosons) and Fermi-Dirac distribution (for fermions), $\mathrm{Im}(\xi(\omega))$ indicates the imaginary part of admittance and $\mathcal{F}^{-1}$ denotes the inverse Fourier transform. In this present scenario the right hand side of the above equation reads
\begin{eqnarray}
\mathcal{F}^{-1}\left[(1+2 n_{n_{\mathrm{B,F}}})\mathrm{Im}(\xi(\omega))\right]&=&\frac{8\alpha^\prime\left(1-\gamma^2\left(\frac{r_{h}}{\tilde{r}_{h}}\right)^d\right)^{\frac{1}{2}}}{\left(\frac{\tilde{r}_{h}}{R}\right)^2\left(1+v^2\gamma^2\left(\frac{r_{h}}{\tilde{r}_{h}}\right)^d\right)^{\frac{1}{2}}\sqrt{f(\tilde{r}_{h})}}\int_{-\infty}^{\infty}\frac{d\omega}{|\omega|}\left(\frac{2}{e^{\beta|\omega|}\pm 1}+1\right)e^{i\omega t}\nonumber\\
&=&G_{\mathrm{sym}}^{\mathrm{B,F}}~.
\end{eqnarray}
The above result implies the fluctuation-dissipation theorem holds in this scenario.
\section{Brownian motion perpendicular to the boost}
In the preceding sections, we have studied the Brownian motion of a heavy particle traversing a strongly coupled plasma with finite temperature and non-zero velocity, employing the gauge/gravity duality framework. The analysis was carried out by considering fluctuations of an open string embedded in a boosted AdS-Schwarzschild black hole background, where the dual description captures the dynamics of the probe particle at the boundary. Specifically, we focused on string fluctuations along the direction of the plasma boost, corresponding to longitudinal motion of the boundary particle. To characterize the stochastic behavior of the particle, we computed the associated diffusion coefficient using two independent holographic approaches. Our results indicate that the presence of the boost leads to a suppression of the diffusion process. The diffusion coefficient decreases, signaling that the boosted motion of the medium impedes the random motion of the boundary particle.\\
In this section, we consider the dynamics of a heavy particle moving in a direction perpendicular to the boost applied in the boundary field theory. Specifically, we assume that the boost is along the $y$-direction, while the heavy particle propagates along the $x$-direction. This configuration is distinct from the parallel case and leads to different physical consequences in both the boundary and bulk descriptions. In the dual gravitational picture, the motion of the particle is typically modeled by a trailing string in an asymptotically AdS geometry, modified by the presence of the boost. The orientation of the particle's trajectory with respect to the boost alters the induced worldsheet metric and consequently the form of the drag force experienced by the particle. Our goal is to compute the diffusion coefficient, admittance and analyze how the transverse motion modifies the standard results obtained in the boosted thermal background.   The influence of the boost introduces anisotropy in the background, and this anisotropy is reflected in the structure of the fluctuation equations and the corresponding response functions.\\
To address the above problem, we consider fluctuations of an open string in a boosted AdS-Schwarzschild background. More precisely, we focus on fluctuations of the string in the direction perpendicular to the boost, which corresponds to the transverse motion of the heavy particle in the boundary theory. Therefore, in this setup, the Nambu-Goto action reads
\begin{eqnarray}\label{perac}
\mathrm{S}_{\mathrm{NG}}^{per}=\frac{1}{2\pi\alpha^\prime}\int d\tau d\sigma \sqrt{-\mathrm{det}\gamma_{ab}^{per}}~
\end{eqnarray}
where $\tau$ and $\sigma$ denote the temporal and spatial coordinate respectively and $\gamma_{ab}^{per}$ represents the induced metric on the two dimensional  string worldsheet. Therefore, the general expression for this worldsheet metric is given by
\begin{eqnarray}
\gamma_{ab}^{per}= g_{\mu\nu}\partial_{a}X^\mu\partial_{b} X^{\nu}~~;~~~a,b=0,1~~;~~ \mu,\nu=0,..., d+1~
\end{eqnarray} 
where $g_{\mu\nu}$ denotes spacetime geometry in the bulk. To proceed, we adopt the static gauge, identifying the worldsheet coordinates with spacetime coordinates as $\tau=t$ and $\sigma=r$. In this gauge, the transverse fluctuations of the string (perpendicular to the boost direction) can be described by 
$X(t,r)$. Thus, the components of the induced metric can be written as
\begin{eqnarray}
\gamma_{tt}^{per}&=&g_{tt}+g_{xx}(\partial_{t}X(r,t))^2\nonumber\\
\gamma_{tr}^{per}&=& \gamma_{rt}^{per}=g_{xx}\partial_{t}X(r,t)\partial_{r}X(r,t)\nonumber\\
\gamma_{rr}^{per}&=&g_{rr}+g_{xx}(\partial_{r}X(r,t))^2~.
\end{eqnarray}
Therefore, the determinant of the world sheet metric reads
\begin{eqnarray}
\mathrm{det}(\gamma_{ab}^{per})&=&\gamma_{tt}^{per}\gamma_{rr}^{per}-(\gamma_{rt}^{per})^2\nonumber\\
&=&g_{tt}g_{rr}+g_{tt}g_{xx}(\partial_{r}X)^2+g_{rr}g_{xx}(\partial_{t}X)^2~.
\end{eqnarray}
Now substituting the above result in eq.(\eqref{perac}), we get the following form of the NG action
\begin{eqnarray}
S_{NG}^{per}&=&-\frac{1}{2\pi\alpha^\prime}\int dt dr~\sqrt{-(g_{tt}g_{rr}+g_{xx}g_{tt}\left(\partial_{r}X(r,t)\right)^2+g_{xx}g_{rr}\left(\partial_{t}X(r,t)\right)^2)}\nonumber\\
&\approx&S_{NG}^{(0)}+\frac{1}{4\pi\alpha^\prime}\int dt dr\left[\frac{g_{xx}(r)\sqrt{|g_{tt}(r)|g_{rr}(r)}}{|g_{tt}(r)|}(\partial_{t}X)^2-\frac{g_{xx}(r)\sqrt{|g_{tt}(r)|g_{rr}(r)}}{g_{rr}(r)}(\partial_{r}X)^2\right]~~.
\end{eqnarray}
The above result includes terms up to second order in time and radial derivatives of 
$X$, that is, in $\partial_{t} X$ and $\partial_{r}X$.  Now varying the above action, we obtain the following equation of motion 
\begin{eqnarray}\label{eom}
\frac{\partial}{\partial r}\left(\frac{g_{xx}(r)\sqrt{|g_{tt}(r)|g_{rr}(r)}}{g_{rr}(r)}\partial_{r}X\right)-\frac{g_{xx}(r)\sqrt{g_{tt}(r)g_{rr}(r)}}{|g_{tt}(r)|}\partial_{t}^2X=0~.
\end{eqnarray}
To get the above equation of motion, we have used the Neumann boundary condition, which is given by following
\begin{eqnarray}
\frac{g_{xx}(r)\sqrt{|g_{tt}(r)|g_{rr}(r)}}{g_{rr}(r)}\partial_{r}X(r,t)|_{\mathrm{boundary}}=0~.
\end{eqnarray}
Now we take the ansatz for $X(r,t)$ of the following form $X(r,t)=f_{\omega}(r)e^{i\omega t}$. Substituting it in eq.\eqref{eom}, we have the following equation for $f_{\omega}(r)$
\begin{eqnarray}\label{eqf}
\frac{\partial}{\partial_r}\left(\left(\frac{r}{R}\right)^4\sqrt{f(r)}\sqrt{1-\gamma^2\left(\frac{r_{h}}{r}\right)^d}\partial_rf_{\omega}(r)\right)+\frac{\omega^2}{\sqrt{f(r)}\sqrt{1-\gamma^2\left(\frac{r_{h}}{r}\right)^d}} f_{\omega}(r)=0~.
\end{eqnarray}
The equation above indicates that obtaining an analytical solution is extremely challenging for a generic background metric. Therefore, in the following sections, we employ the standard patching method to construct an approximate analytical solution.
\subsection{General solution}
\noindent As discussed earlier, an exact analytical solution to the above equation is not feasible. To proceed with the patching method and construct an approximate analytical solution, we consider the following cases: (a) near horizon region (IR region), (b) hydrodynamic limit ($\omega\rightarrow 0$), (c) UV region ($r\rightarrow\infty$).
\subsubsection{Solution in the near horizon region}
In this section, we obtain the solution of eq.\eqref{eqf} in the near horizon region. Let us denote the solution of this equation by $f_{\omega}^{A}(r)$. To obtain an analytical solution of this equation in the near horizon region, we have to recast the equation given in eq.\eqref{eqf} in terms of the tortoise coordinate
\begin{eqnarray}
\frac{dr_{*}}{dr}=\frac{\gamma R^2}{r^2 f(r)}~.
\end{eqnarray}
Therfore, using the above result one can rewrite the eq.\eqref{eqf} in the following form
\begin{eqnarray}
	\frac{d}{dr_{*}}\left[\left(\frac{r}{R}\right)^2\frac{\sqrt{1-\gamma^2\left(\frac{r_{h}}{r}\right)^d}}{\sqrt{f(r)}}\frac{df_{\omega}^{A}(r)}{dr_{*}}\right]+\frac{\omega^2}{\gamma^2}\frac{\left(\frac{r}{R}\right)^2\sqrt{f(r)}}{\sqrt{1-\gamma^2\left(\frac{r_{h}}{r}\right)^d}}f_{\omega}^{A}(r)=0~.
	\end{eqnarray}
To proceed further, let us write down the above equation in the following form
\begin{eqnarray}\label{EQ10}
\frac{d}{dr_{*}}\left[\tilde{\alpha}(r)\frac{df_{\omega}^{A}(r)}{dr_{*}}\right]+\frac{\omega^2}{\gamma^2}\tilde{\delta}(r)f_{\omega}^{A}(r)=0~
\end{eqnarray}
where the expression of $\tilde{\alpha}(r)$ and $\tilde{\delta}(r)$ are given by followig
\begin{eqnarray}
\tilde{\alpha}(r)&=&\left(\frac{r}{R}\right)^2\frac{\sqrt{1-\gamma^2\left(\frac{r_{h}}{r}\right)^d}}{\sqrt{f(r)}}~;~\tilde{\delta}(r)=\frac{\left(\frac{r}{R}\right)^2\sqrt{f(r)}}{\sqrt{1-\gamma^2\left(\frac{r_{h}}{r}\right)^d}}
\end{eqnarray}
One can further simplify eq.\eqref{EQ10} in the following form
\begin{eqnarray}\label{EQ11}
\frac{d^2f_{\omega}^A(r)}{dr_*^2}+\frac{d\log(\tilde{\alpha}(r))}{dr_*}\frac{d f_{\omega}^{A}(r)}{dr_*}+\frac{\omega^2}{\gamma^2}\frac{\tilde{\delta}(r)}{\tilde{\alpha}(r)}f_{\omega}^{A}(r)=0~~~.
\end{eqnarray}
Now let us consider the following ansatz for $f_{\omega}^{A}$
\begin{eqnarray}
f_{\omega}^{A}(r)=e^{-\frac{B(r_*)}{2}}\phi_{\omega}(r)
\end{eqnarray}
Upon substituting the above ansatz in eq.\eqref{EQ11} we get
\begin{eqnarray}\label{scho1}
\frac{d^2\phi_{\omega}}{dr_{*}^2}+\left(\frac{\omega^2}{\gamma^2}\frac{\tilde{\delta}(r)}{\tilde{\alpha}(r)}-\tilde{V}(r)\right)\phi_{\omega}=0~.
\end{eqnarray}
The above result shows that the differential equation contains no term involving the first derivative of $\phi_{\omega}(r)$. This is because we have made the following identification
\begin{eqnarray}
B=\log(\tilde{\alpha}(r))~.
\end{eqnarray}
Using this identification, we can express $V(r)$ and $f_{\omega}^{A}(r)$ as
\begin{eqnarray}
\tilde{V}(r)&=&\frac{1}{2}\frac{r^2 f(r)}{\gamma R^2}\frac{d}{dr}\left(\frac{r^2f(r)}{\gamma R^2}\frac{d \log(\tilde{\alpha}(r))}{dr}\right)+\frac{1}{4}\frac{r^4f^{2}(r)}{\gamma^2 r^4}\left(\frac{d \log(\tilde{\alpha}(r))}{dr}\right)^2\\
f_{\omega}^{A}(r)&=&\frac{\phi_{\omega}(r)}{\sqrt{\tilde{\alpha}(r)}}
\end{eqnarray}
A detailed derivation of the above equation for diagonal bulk metric can be found in \cite{Caldeira:2020rir,Caldeira:2020sot,Caldeira:2021izy,Caldeira:2022gfo}. Furthermore, in the near horizon region, that is, $r\approx \tilde{r}_{h}=r_{h}=1+2\epsilon$ \cite{deBoer:2008gu}, $V(r) \rightarrow 0$, the Schr\"{o}dinger like equation (given in eq.\eqref{scho1})  can be written as
\begin{eqnarray}\label{scho2}
\frac{d^2\phi_{\omega}}{dr_{*}^2}+\frac{\omega^2}{\gamma^2}F(\tilde{r}_{h})\phi_{\omega}=0~.
\end{eqnarray}
where the expression of $F(\tilde{r}_{h})$ is given as 
\begin{eqnarray}
F(\tilde{r}_{h})=\frac{f(\tilde{r}_{h})}{\left[1-\gamma^2\left(\frac{r_{h}}{\tilde{r}_{h}}\right)^d\right]}
\end{eqnarray}\\
The general solution of the above equation can be written as 
\begin{eqnarray}
\phi_{\omega}=\tilde{A}_{1}e^{-i \frac{\omega}{\gamma}\sqrt{F(\tilde{r}_{h})}~ r_{*}}+\tilde{A}_{2}e^{i \frac{\omega}{\gamma}\sqrt{F(\tilde{r}_{h})}~r_{*}}~.
\end{eqnarray}
To proceed further, we focus exclusively on the ingoing solution, as it is the physically relevant mode near the horizon, corresponding to waves falling into the black hole. Therefore, we choose the constant $\tilde{A}_{2}$ to be zero, that is, $\tilde{A}_{2}=0$. Keeping this fact in mind, the solution of the above equation (given in eq.\eqref{scho2}) can be written as
\begin{eqnarray}
f_{\omega}^{A}(r)=\frac{\tilde{A}_{1}e^{-i \frac{\omega}{\gamma}\sqrt{F(\tilde{r}_{h})}~ r_{*}}}{\sqrt{\tilde{\alpha}(\tilde{r}_{h})}}~.
\end{eqnarray}
Before proceeding further, we comment on our choice of taking the near-horizon limit 
$r \to \tilde r_h$ instead of $r \to r_h$. The above equation indicates that the solution 
diverges as $r \to r_h$. This divergence originates from $\tilde{\alpha}(r) \propto f(r)$, where 
$f(r)$ is the blackening function that vanishes at the horizon. This feature is specific 
to non-diagonal bulk metrics and is absent for diagonal metrics.\\
However, in the small-frequency limit, the above result simplifies to the following form
\begin{eqnarray}\label{fir}
f_{\omega}^{A}(r)&\approx&\frac{\tilde{A}_{1}}{\sqrt{\tilde{\alpha}(\tilde{r}_{h})}}\left(1-i\frac{\omega}{\gamma}\sqrt{F(\tilde{r}_{h})}r_{*}\right)~.
\end{eqnarray}
Now substituting the expression of tortoise coordinate in the above equation we get
\begin{eqnarray}
f_{\omega}^{A}(r)=\frac{\tilde{A}_{1}\Bigg[1-i\omega\sqrt{F(\tilde{r}_{h})}R^2\int dr~\frac{1}{r^2 f(r)}\Bigg]}{\sqrt{\tilde{\alpha}(\tilde{r}_{h})}}~.
\end{eqnarray}
Proceeding as before, the near-horizon solution is obtained by evaluating the integral, yielding
\begin{eqnarray}\label{EQ12}
f_{\omega}^{A}(r)\approx\frac{\tilde{A}_1}{\sqrt{\tilde{\alpha}(\tilde{r}_{h})}}\left[1-\frac{i\omega R^2\sqrt{F(\tilde{r}_{h})}}{\tilde{r}_{h}^2f^{\prime}(\tilde{r}_{h})}\log\left(\frac{r}{\tilde{r}_{h}}-1\right)\right]~~.
\end{eqnarray}
\subsubsection{Solution in the hydrodynamic regime}
\noindent In this section, we obtain the solution of eq.\eqref{eqf} in  the hydrodynamic regime. In this regime we can neglect the second term of eq.\eqref{eqf}, which contains $\omega^2$. Therefore, in this regime eq.\eqref{eqf} reduces to the following form
\begin{eqnarray}
\frac{\partial}{\partial r}\left(\left(\frac{r}{R}\right)^4\sqrt{f(r)}\sqrt{1-\gamma^2\left(\frac{r_{h}}{r}\right)^d}\partial_rf_{\omega}^{B}(r)\right)=0~.\nonumber\\
\end{eqnarray}
Therefore, the general solution of the equation can be written as
\begin{eqnarray}
f_{\omega}^{B}(r)=\tilde{B}_{1}(\omega)\int^{r} dr^\prime \frac{\left(\frac{R}{r^{\prime}}\right)^4}{\sqrt{f(r^\prime)}\sqrt{1-\gamma^2\left(\frac{r_{h}}{r^\prime}\right)^d}}
+ \tilde{B}_{2}(\omega)~
\end{eqnarray}
where $\tilde{B}_{1}(\omega)$ and $\tilde{B}_{2}(\omega)$ are constants depending only on the frequency $\omega$. To proceed, we now consider the near-horizon approximation. In this limit, the above integral  can be approximated in the following simplified form
\begin{eqnarray}
f_{\omega}^{B}(r)|_{\mathrm{IR}}&\approx&\frac{\tilde{B}_{1}(\omega)\left(\frac{R}{\tilde{r}_{h}}\right)^4}{\sqrt{1-\gamma^2\left(\frac{r_{h}}{\tilde{r}_{h}}\right)^d}}\int ^{r}dr^\prime \frac{1}{\sqrt{f(r^\prime)}} + \tilde{B}_{2}(\omega)\nonumber\\
&\approx&\frac{\tilde{B}_{1}(\omega)\left(\frac{R}{\tilde{r}_{h}}\right)^4\sqrt{f(\tilde{r}_{h})}}{\sqrt{1-\gamma^2\left(\frac{r_{h}}{\tilde{r}_{h}}\right)^d}f^{\prime}(\tilde{r}_{h})}\log\left(\frac{r}{\tilde{r}_{h}}-1\right) + \tilde{B}_{2}(\omega)~.
\end{eqnarray}
By comparing the above result with the expression given in eq.~\eqref{EQ12}, we can determine $\tilde{B}_{1}(\omega)$ and $\tilde{B}_{2}(\omega)$ in terms of $\tilde{A}_{1}(\omega)$. This leads to the following relations
By comparing the above result with the expression given in eq.~\eqref{EQ12}, we can determine $\tilde{B}_{1}(\omega)$ and $\tilde{B}_{2}(\omega)$ in terms of $\tilde{A}_{1}(\omega)$. This leads to the following relations
\begin{eqnarray}\label{B1}
\tilde{B}_{1}(\omega)&=&\frac{\tilde{A}_{1}(\omega)}{\sqrt{\tilde{\alpha}(\tilde{r}_{h})}}~;~\tilde{B}_{1}(\omega)=-i\omega\tilde{A}_{1}(\omega)\left(\frac{\tilde{r}_{h}}{R}\right)\frac{f^{\frac{1}{4}}(\tilde{r}_{h})}{\left[1-\gamma^2\left(\frac{r_{h}}{\tilde{r}_{h}}\right)^d\right]^{\frac{1}{4}}}~.
\end{eqnarray}
One can also determine $\tilde{A}_{1}(\omega)$ by the normlisation method, which we have discussed earlier \subsubsection{Solution in the UV domain}
In this section, we obtain the solution of eq.~\eqref{eqf} in the ultraviolet (UV) region, that is, in the limit \( r \to \infty \). In this regime the equation of motion for string fluctuation reduces to
\begin{eqnarray}
\frac{\partial}{\partial r}\left(\left(\frac{r}{R}\right)^4\sqrt{f(r)}\sqrt{1-\gamma^2\left(\frac{r_{h}}{r}\right)^d}\partial_rf_{\omega}^{C}(r)\right)=0~.
\end{eqnarray}
The above equation is very similar to that of in the hydrodynamic regime. Therefore, the soultion of the above equation can be written as 
\begin{eqnarray}\label{fuv}
f_{\omega}^{C}(r)=\frac{\tilde{A}_{1}f^{\frac{1}{4}}(\tilde{r}_{h})}{\left(\frac{\tilde{r}_{h}}{R}\right)\left[1-\gamma^2\left(\frac{r_{h}}{\tilde{r}_{h}}\right)^d\right]^{\frac{1}{4}}}\left[1-i\omega\left(\frac{\tilde{r}_{h}}{R}\right)^2\int_{\tilde{r}_{h}}^{r}dr^\prime \frac{\left(\frac{R}{r^{\prime}}\right)^4}{\sqrt{f(r^\prime)}\sqrt{1-\gamma^2\left(\frac{r_{h}}{r^\prime}\right)^d}}\right]
\end{eqnarray}
To get the above expression, we have used the results given in eq.\eqref{B1}. The analytical solution obtained in the UV limit will serve as a fundamental input in the computation of the two-point thermal correlator, and consequently, in the determination of the diffusion coefficient via linear response theory. In the subsequent analysis, we utilize the previously obtained solution to compute the diffusion coefficient through two complementary approaches, namely, linear response theory and the evaluation of the two-point thermal correlator.
\subsection{Computation of admittance and diffusion coefficient}
\begin{figure}[!h]
\begin{minipage}[t]{0.48\textwidth}
\centering\includegraphics[width=\textwidth]{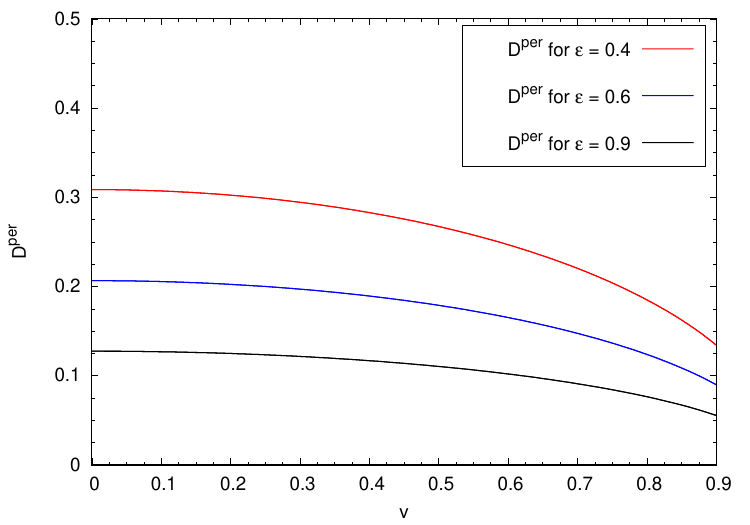}\\
{\footnotesize variation of $D^{per}$ w.r.t $v$}
\end{minipage}\hfill
\begin{minipage}[t]{0.48\textwidth}
\centering\includegraphics[width=\textwidth]{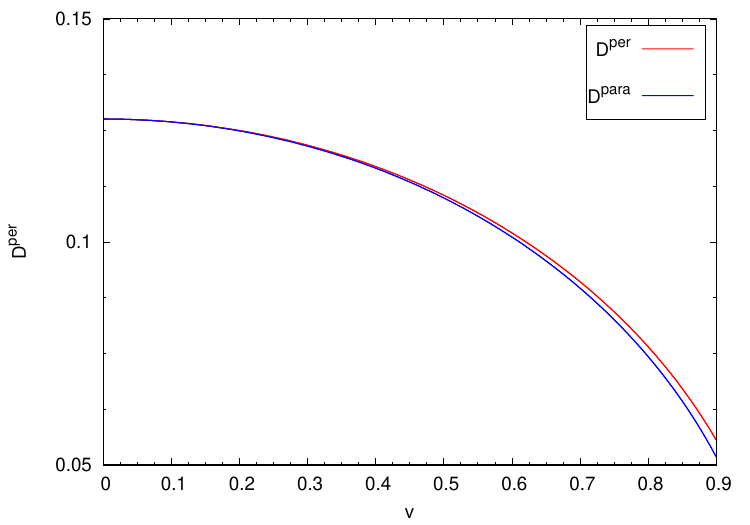}\\
{\footnotesize Comparison between $D^{para}$ and $D^{per}$ }
\end{minipage}
\caption{The left panel of the above Figure represents the variation of $D^{per}$ with respect to the boost velocity $v$. To do this plot we have set $d=4,\alpha=10, r_{h}=20$. On the other hand, the right panel of the Figure presents a comparison between the two diffusion coefficients, $D^{\mathrm{per}}$ and $D^{\mathrm{para}}$. The black and red lines represent the diffusion coefficients  parallel and perpendicular to the direction of the boost respectively. Again to do this plot we have set $d=4,\alpha=10, r_{h}=20, v=0.9$.}
\label{Diff}
\end{figure}

\noindent Keeping the above discussion in mind, we now proceed to compute the admittance and, subsequently, the diffusion coefficient for a Brownian particle moving through a boosted thermal plasma along the perpendicular to the direction of boost. Admittance can be defined as the linear response of a system to an external perturbation.  Therefore, we need to introduce a small external force acting on the boundary particle. In this scenario we introduce an external electric field $A_{\mu}$ to the UV brane in the arbitrary direction $x^{i}$. This external electric field does not alter the bulk dynamics of the string, but it exerts a force on the string's endpoints located on the ultraviolet (UV) brane.
Therefore, in the presence of an external field, the Nambu-Goto action can be written as
\begin{eqnarray}
S^{per}_{NG}&\approx& S_{NG}^{(0)}+\frac{1}{4\pi\alpha^\prime}\int dt dr\Bigg[\frac{g_{xx}(r)\sqrt{|g_{tt}(r)|g_{rr}(r)}}{|g_{tt}(r)|}(\partial_{t}X)^2-\frac{g_{xx}(r)\sqrt{|g_{tt}(r)|g_{rr}(r)}}{g_{rr}(r)}(\partial_{r}X)^2\Bigg]\nonumber\\
&+&\int dt \left(A_{t}+\vec{A}.\vec{\dot{x}}\right)|_{r=r_{b}}~.
\end{eqnarray}
Varying the total action, which includes the coupling to the external electric field, yields a modified boundary condition. 
The modified boundary condition in the presence of the external electromagnetic field reads
\begin{eqnarray}
F(t)=\frac{1}{2\pi\alpha^\prime}\left(\frac{g_{xx}(r)\sqrt{|g_{tt}(r)|g_{rr}(r)}}{g_{rr}(r)}\partial_r X(r,t)\right)_{r=r_{b}}
\end{eqnarray}
where $F(t)=\partial_{x}A_{t}-\partial_{t}A_{x}$. Now substituting the expression of $X(r,t)=f_{\omega}^{C}(r)e^{i\omega t}$ in the above result, we get
\begin{eqnarray}
F(\omega)\approx-\frac{i\omega \tilde{A}_{1}(\omega)\left(\frac{\tilde{r}_{h}}{R}\right)}{2\pi \alpha^\prime}\frac{\left(1-\left(\frac{r_{h}}{\tilde{r}_{h}}\right)^d\right)^{\frac{1}{4}}}{\left(1-\gamma^2\left(\frac{r_{h}}{\tilde{r}_{h}}\right)^d\right)^{\frac{1}{4}}}~.
\end{eqnarray}
\noindent Therefore, in this present scenario the admittance can be computed by substituting the expression of $F(\omega)$ and the solution of eq.\eqref{eqf} in the UV domain (given in eq.\eqref{fuv}) in the formula given by eq.\eqref{adm}. This results in
\begin{eqnarray}
\xi^{per}(\omega)=2\pi\alpha^\prime\frac{\frac{\tilde{A}_{1}f^{\frac{1}{4}}(\tilde{r}_{h})}{\left(\frac{\tilde{r}_{h}}{R}\right)\left[1-\gamma^2\left(\frac{r_{h}}{\tilde{r}_{h}}\right)^d\right]^{\frac{1}{4}}}\left[1-i\omega\left(\frac{\tilde{r}_{h}}{R}\right)^2\int_{\tilde{r}_{h}}^{r}dr^\prime \frac{\left(\frac{R}{r^{\prime}}\right)^4}{\sqrt{f(r^\prime)}\sqrt{1-\gamma^2\left(\frac{r_{h}}{r^\prime}\right)^d}}\right]}{i\omega \tilde{A}_{1}(\omega)\left(\frac{\tilde{r}_{h}}{R}\right)\frac{\left(1-\left(\frac{r_{h}}{\tilde{r}_{h}}\right)^d\right)^{\frac{1}{4}}}{\left(1-\gamma^2\left(\frac{r_{h}}{\tilde{r}_{h}}\right)^d\right)^{\frac{1}{4}}}}~.
\end{eqnarray}	
Thus, the imaginary part of the admittance can be written as
\begin{eqnarray}\label{adm1}
\mathrm{Im}(\xi^{per}(\omega))=\frac{2\pi\alpha^\prime}{\omega\left(\frac{\tilde{r}_{h}}{R}\right)^2}~.
\end{eqnarray}
Before proceeding further, it is important to observe that the functional form of the admittance in the present setup coincides with that obtained in the case of an unboosted Schwarzschild black hole. This indicates that for a Brownian particle undergoing motion transverse to the boost direction, the admittance remains invariant under the influence of the boost. Now keeping this above result in mind, we now compute the diffusion coefficient  by the following formula \cite{PhysRevLett.110.061602}
\begin{equation}
D^{per}=\frac{1}{\beta}~\mathop{\mathrm{lim}}_{\omega\rightarrow 0}\left(-i \omega\xi^{per}(\omega)\right)~.
\end{equation}
We get an explicit form of diffusion coefficient by substituting the expression of $	\mathrm{Im}(\xi^{per}(\omega)$ and $\beta$ in the above result. This yields
\begin{eqnarray}\label{Dper}
D^{per}=\frac{2\pi\alpha^\prime T}{\left(\frac{\tilde{r}_{h}}{R}\right)^2}~.
\end{eqnarray}
This result indicates that in the limit \( v \to 0 \), the expression for the diffusion coefficient consistently reduces to its counterpart in the unboosted Schwarzschild background, as expected. The above result implies that the diffusion coefficient in the perpendicular case is samaller than that in unboosted scenario.
This implies that the diffusion process becomes slower in the presence of a boost. We have represented this result in the left panel of Fig.\eqref{Diff}. It is further observed that diffusion along the direction of boost experiences greater suppression relative to the transverse direction, indicating anisotropic transport behavior induced by the boost. This implies that the diffusion coefficient along the boost is smaller than that of perpendicular to the boost for same value of boost parameter, that is, $D^{para}\le D^{per}$\footnote{The equality sign holds for $v=0$ scenario.}.
\noindent This observation is represented graphically in the right panel of Fig.\eqref{Diff}. In this figure we have plotted both the $D^{para}$ and $D^{per}$ as functions of the boost parameter. 
The red and blue curves represent the diffusion coefficients perpendicular and parallel to the boost direction, respectively. It is also evident from the expression of the diffusion coefficient that, in this scenario, the IR cutoff does not play a crucial role as it does in the parallel direction.
\subsection{Computation of correlation function and diffusion coefficient}
In this section, we evaluate the diffusion coefficient using the thermal two-point correlation function of the string endpoint at the UV brane. Specifically, the diffusion coefficient is obtained from the mean square displacement, which characterizes the variance of the particle's stochastic trajectory due to thermal fluctuations. The computation of the thermal two-point correlator relies on the previously derived solution for the string fluctuations in the UV and near-horizon limits.\\
The general solution of eq.\eqref{eqf} in the near horizon region can be written as
\begin{eqnarray}
f_{\omega}^{\mathrm{IR}}(r)=\tilde{A}\left[e^{i \frac{\omega}{\gamma}\sqrt{F(\tilde{r}_{h})} r_{*}}+\tilde{B}~e^{-i \frac{\omega}{\gamma}\sqrt{F(\tilde{r}_{h})} r_{*}}\right]~.
\end{eqnarray}
On the other hand, in the UV region the solution of eq.\eqref{eqf} reads
\begin{eqnarray}\label{UV}
f_{\omega}^{\mathrm{UV}}(r)&=&\frac{\tilde{A}_{1}f^{\frac{1}{4}}(\tilde{r}_{h})}{\left(\frac{\tilde{r}_{h}}{R}\right)\left[1-\gamma^2\left(\frac{r_{h}}{\tilde{r}_{h}}\right)^d\right]^{\frac{1}{4}}}\Bigg[\left(1-i\omega\left(\frac{\tilde{r}_{h}}{R}\right)^2\int_{\tilde{r}_{h}}^{r}dr^\prime \frac{\left(\frac{R}{r^{\prime}}\right)^4}{\sqrt{f(r^\prime)}\sqrt{1-\gamma^2\left(\frac{r_{h}}{r^\prime}\right)^d}}\right)\nonumber\\
&+&\tilde{B}\left(1+i\omega\left(\frac{\tilde{r}_{h}}{R}\right)^2\int_{\tilde{r}_{h}}^{r}dr^\prime \frac{\left(\frac{R}{r^{\prime}}\right)^4}{\sqrt{f(r^\prime)}\sqrt{1-\gamma^2\left(\frac{r_{h}}{r^\prime}\right)^d}}\right)\Bigg]~.
\end{eqnarray}
Here, we have assumed that the constants \(\tilde{A}\) and \(\tilde{B}\) take the same values in both domains under consideration.\\
We now determine the constants \(\tilde{A}\) and \(\tilde{B}\) by following a procedure analogous to that outlined in the earlier section (Section.\eqref{AB}).
We can find $\tilde{B}$ by applying the Neumann boundary condition at the boundary $r=r_b$. We can easily show that at the leading order in $\omega$, the constant $\tilde{B}$ simplifies to unity, that is ,$\tilde{B}=1$. However, it can be easily shown that, for higher order in $\omega$, the constant $\tilde{B}$ is a pure phase factor of the form $\tilde{B}=e^{i\theta \omega}$.\\
On the other hand, applying the Neumann boundary condition near the black hole horizon, we get the following result \begin{eqnarray}
\tilde{B}=e^{2i \frac{\omega}{\gamma}\sqrt{F(\tilde{r}_{h})}}|_{r=\tilde{r}_{h}(1+\epsilon)}~~.
\end{eqnarray}
One can get the above result by using the IR solution. To obtain the explicit form of \(\tilde{B}\), we substitute the expression for the tortoise coordinate in the near-horizon limit into the above result. This yields
\begin{eqnarray}
\tilde{B}=\exp(-\frac{2i\omega R^2\sqrt{F(\tilde{r}_{h})}}{f^{\prime}(\tilde{r}_{h}) \tilde{r}_{h}^2}\log(\frac{1}{\epsilon}))~.
\end{eqnarray}
The result once again confirms that \(\tilde{B}\) is a pure phase factor, depending only on the frequency \(\omega\). Before proceeding further, we note that the Neumann boundary condition is imposed in the near-horizon region. Imposing the Dirichlet boundary condition at the horizon leads to a trivial solution, eliminating non-trivial IR modes. Hence, the Neumann boundary condition is required in the near-horizon sector, as also discussed in \cite{deBoer:2008gu}. Before proceeding further, we would like to make few comments. As previously mentioned, since \(\epsilon \ll 1\), it follows that the frequency spectrum is discrete rather than continuous. 
	The discrete nature of the frequency spectrum is characterized by the spacing \(\Delta \omega\)
	\begin{eqnarray}
	\Delta\omega&=&\frac{\pi \left(\frac{\tilde{r}_{h}}{R}\right)^2f^{\prime}(\tilde{r}_{h})}{\log(\frac{1}{\epsilon})\sqrt{F(\tilde{r}_{h})}}=\frac{\pi \left(\frac{\tilde{r}_{h}}{R}\right)^2f^{\prime}(\tilde{r}_{h})\left(1-\gamma^2\left(\frac{r_{h}}{\tilde{r}_{h}}\right)^d\right)^{\frac{1}{2}}}{\log(\frac{1}{\epsilon})\left(1-\left(\frac{r_{h}}{\tilde{r}_{h}}\right)^d\right)^{\frac{1}{2}}}~.
	\end{eqnarray}
	Therefore, the density of states reads
	\begin{eqnarray}
	\mathcal{D}(\omega)=\frac{\log(\frac{1}{\epsilon})\left(1-\left(\frac{r_{h}}{\tilde{r}_{h}}\right)^d\right)^{\frac{1}{2}}}{\pi \left(\frac{\tilde{r}_{h}}{R}\right)^2f^{\prime}(\tilde{r}_{h})\left(1-\gamma^2\left(\frac{r_{h}}{\tilde{r}_{h}}\right)^d\right)^{\frac{1}{2}}}~.
	\end{eqnarray}
An analytical expression for \(\tilde{A}\) can also be obtained using the standard normalization procedure. 
This method has been discussed in detail in Section\eqref{AB}. Applying the same technique here yields the following expression for \(\tilde{A}\)
\begin{eqnarray}
|\tilde{A}|^{2}&\approx&\frac{\pi \alpha^\prime f^{\prime}(\tilde{r}_{h})\left(1-\gamma^2\left(\frac{r_{h}}{\tilde{r}_{h}}\right)^d\right)^{\frac{1}{2}}}{\omega\sqrt{f(\tilde{r}_{h})}\log(\frac{1}{\epsilon})}
=\frac{\alpha^\prime}{\left(\frac{\tilde{r}_{h}}{R}\right)^2}\frac{\Delta \omega}{\omega}~.
\end{eqnarray}
\subsubsection{Computation of correlation function and mean square displacement}
\noindent In this section, we compute the correlation functions and the mean square displacement of the string endpoints on the UV brane, under the assumption of zero chemical potential. To begin, we consider the density matrix for a canonical ensemble, which is given by eq.\eqref{den}. To proceed further, we express the solution of eq.\eqref{eqf} in terms of the creation (\(a_{\omega}^{\dagger}\)) and annihilation (\(a_{\omega}\)) operators as follows
\begin{eqnarray}
X(r,t)=\sum_{\omega>0}\left(a_{\omega}f_{\omega}^{\mathrm{UV}}(r)e^{-i\omega t}+a_{\omega}^{\dagger}(f_{\omega}^{\mathrm{UV}}(r))^{*}e^{-i\omega t}\right)~.
\end{eqnarray}
In this expression, \(f_{\omega}^{\mathrm{UV}}(r)\) corresponds to the solution of eq.~\eqref{eqf} in the UV regime, whose explicit form is provided in eq.~\eqref{UV}.
\noindent We now define the mean square displacement \( s^{2}(t) \) as follows
\begin{eqnarray}\label{meper}
(s^{2}(t))^{per}&\equiv&\langle[x(t)-x(0)]^2\rangle=\langle[X(t,r_{b})-X(0,r_{b})]^2\rangle
\end{eqnarray}
where $x(t)=X(t,r_{b})$ represents the position of the external particle in the boundary theory.
\noindent In order to derive an explicit expression for the transverse mean square displacement \( (s^{2}(t))^{\perp} \), we first evaluate the following thermal expectation value
\begin{eqnarray}\label{expv1}
\langle x(t_{1})x(t_{2})\rangle&=&\langle X(t_{1},r_{b})X(t_{2},r_{b})\rangle\nonumber\\
&=&\sum_{\omega,\omega^\prime>0}\Bigg(\langle a_{\omega}^{\dagger}a_{\omega^{\prime}}\rangle_{\rho_{0}}(f_{\omega}^{\mathrm{UV}*}(r_{b})f_{\omega^{\prime}}^{\mathrm{UV}}(r_{b}))e^{i\omega t_{1}-i\omega^{\prime}t_{2}}+\langle a_{\omega}a_{\omega^{\prime}}^{\dagger}\rangle_{\rho_{0}}(f_{\omega}^{\mathrm{UV}}(r_{b})f_{\omega^{\prime}}^{\mathrm{UV}*}(r_{b}))e^{-i\omega t_{1}+i\omega^{\prime}t_{2}}\Bigg)\nonumber\\	&=&\sum_{\omega>0}\left[|f_{\omega}^{\mathrm{UV}}(r_{b})|^2\frac{2\cos(\omega(t_{1}-t_{2}))}{e^{\beta\omega}\pm1}+|f_{\omega}^{\mathrm{UV}}(r_{b})|^2 e^{i\omega(t_{2}-t_{1})}\right]~.
\end{eqnarray}
In the second line of the above equation we have used the results given in eq.\eqref{expv}.
Furthermore, setting \( t_{1} = t \) and \( t_{2} = 0 \) in the above expression, we obtain
\begin{eqnarray}
\langle x(t)x(0)\rangle=\sum_{\omega>0}|f_{\omega}^{\mathrm{UV}}(r_{b})|^2\left(\frac{2 \cos(\omega t)}{e^{\beta\omega}\pm1}+e^{-i\omega t}\right)~.
\end{eqnarray}
Now substituting the expression of $f_{\omega}^{\mathrm{UV}}(r_{b})$ (given in eq.\eqref{UV}) in the above result, we get
\begin{eqnarray}\label{x}
\langle x(t)x(0)\rangle=\sum_{\omega>0}\frac{\pi \alpha^\prime f^{\prime}(\tilde{r}_{h})\left(1-\gamma^2\left(\frac{r_{h}}{\tilde{r}_{h}}\right)^d\right)^{\frac{1}{2}}}{\omega\sqrt{f(\tilde{r}_{h})}\log(\frac{1}{\epsilon})}\left(\frac{2 \cos(\omega t)}{e^{\beta\omega}\pm1}+e^{-i\omega t}\right)~.
\end{eqnarray}
It should be emphasized that the derivation of the above expression involves truncating the expansion at \(\mathcal{O}\left(\frac{1}{\omega^2}\right)\). This approximation is appropriate since our interest lies in the low-frequency limit. To proceed further, we approximate the discrete sum over frequencies by a definite integral using the standard continuum limit:
\[
\sum_{n}\Delta\omega \longrightarrow \int d\omega\Leftrightarrow\sum_{\omega>0}\frac{\pi \alpha^\prime f^{\prime}(\tilde{r}_{h})\left(1-\gamma^2\left(\frac{r_{h}}{\tilde{r}_{h}}\right)^d\right)^{\frac{1}{2}}}{\omega\sqrt{f(\tilde{r}_{h})}\log(\frac{1}{\epsilon})}\rightarrow\int_{0}^{\infty} d\omega~
\]
where \(\Delta \omega\) denotes the spacing between adjacent frequency modes. This substitution is valid in the limit where the spectrum becomes quasi-continuous.\\
Keeping this fact in mind, the integral given in Eq.~\eqref{x} can be rewritten as
\begin{eqnarray}\label{cor1}
\langle x(t)x(0)\rangle&=&\frac{\alpha^\prime}{\left(\frac{\tilde{r}_{h}}{R}\right)^2}\int_{0}^{\infty}\frac{d \omega}{\omega}\left(\frac{2 \cos(\omega t)}{e^{\beta\omega}\pm1}+e^{-i\omega t}\right)\nonumber\\&=&\langle x(0)x(t)\rangle^*~.
\end{eqnarray}
On the other hand, we can obtain $\langle x(t)x(t)\rangle $ and $\langle x(0)x(0)\rangle $ by setting $t_{1}=t_{2}=t$ and $t_{1}=t_{2}=0$ in eq.\eqref{expv1}. This results in
\begin{eqnarray}\label{cor2}
\langle x(t)x(t)\rangle&=&\frac{\alpha^\prime}{\left(\frac{\tilde{r}_{h}}{R}\right)^2}\int_{0}^{\infty}\frac{d \omega}{\omega}\left(\frac{2}{e^{\beta\omega}\pm1}+1\right)=\langle x(0)x(0)\rangle~.
\end{eqnarray}
Based on the preceding results, we now proceed to evaluate the mean square displacement using eq.\eqref{meper}. In particular,  \((s^{2}(t))^{\perp}\) is obtained by substituting the expressions derived in eqs.\eqref{cor1} and \eqref{cor2} into eq.\eqref{meper}. It should be noted that the mean square displacement obtained here is a divergent quantity. This motivates us to define the regularized mean square displacement as follows
\begin{eqnarray}
s_{reg}^{2}(t)|^{perp}&\equiv&\langle:[x(t)-x(0)]^2:\rangle\nonumber\\&=&\langle:[X(t,r_{b})-X(0,r_{b})]^2:\rangle~
\end{eqnarray}
where $:[...]:$ represents the normal ordering. The above expression can be simplified further as follows
\begin{eqnarray}\label{sper}
s_{reg}^{2}(t)|^{perp}=2\langle:x^{2}(t):\rangle-2\langle:x(t)x(0):\rangle~.
\end{eqnarray}
\noindent Now to get an explicit result for the regularized mean square displacement we need to substitute the expressions of $langle:x^{2}(t):\rangle $ and $ \langle :x(t)x(0):\rangle$ in the above equation. One can compute the expressions of $\langle:x^{2}(t):\rangle$ and $\langle:x(t)x(0):\rangle$ by using the eq.\eqref{expv1}.  This yields
\begin{eqnarray}
\langle:x(t)x(0):\rangle&=&\frac{\alpha^\prime}{\left(\frac{\tilde{r}_{h}}{R}\right)^2}\int_{0}^{\infty}\frac{d \omega}{\omega}\frac{2 \cos(\omega t)}{e^{\beta\omega}\pm1}=\langle:x(0)x(t):\rangle^*\nonumber\\
\langle:x(t)x(t):\rangle&=&\frac{\alpha^\prime}{\left(\frac{\tilde{r}_{h}}{R}\right)^2}\int_{0}^{\infty}\frac{d \omega}{\omega}\frac{2}{e^{\beta\omega}\pm1}=\langle:x(0)x(0):\rangle~.
\end{eqnarray}
Substituting the above results, in eq.\eqref{sper} we get
\begin{eqnarray}\label{sPer}
s_{reg}^{2}(t)|^{perp}=\frac{8\alpha^\prime}{\left(\frac{\tilde{r}_{h}}{R}\right)^2}\int_{0}^{\infty}\frac{d \omega}{\omega}\frac{ \sin^2(\frac{\omega t}{2})}{e^{\beta\omega}\pm1}~.
\end{eqnarray}
With the general result for the regularized mean square displacement established, we proceed to compute this quantity in several distinct scenarios in the following section.
\subsubsection{Computation of regularized mean square displacement (RMSD) in different cases }
\noindent In this section, we now calculate the RMSD for different scenarios.
The regularized mean square displacement is computed for both bosonic and fermionic cases. In the bosonic case, it is derived by considering negative sign in the denominator of the general expression obtained earlier in eq.\eqref{sPer}.
Therefore, the expression of RMSD for bosonic particles reads
\begin{eqnarray}
(s^{2}(t))^{per}_{\mathrm{Boson}}&=&\frac{8\alpha^\prime}{\left(\frac{\tilde{r}_{h}}{R}\right)^2}\int_{0}^{\infty}\frac{d \omega}{\omega}\frac{ \sin^2(\frac{\omega t}{2})}{e^{\beta\omega}-1}\nonumber\\
&=&\frac{2\alpha^\prime}{\left(\frac{\tilde{r}_{h}}{R}\right)^2}\log(\frac{\sinh(\frac{t\pi}{\beta})}{\frac{t\pi}{\beta}})~.
\end{eqnarray}
Keeping this  general expression of RMSD in mind, we now proceed to find the behavior of RMSD in two asymptotic time regimes. Specifically, we consider: (i) the \textit{ballistic regime}, which corresponds to very short times such that \( t \ll \beta \), where inertial effects dominate; and (ii) the \textit{diffusive regime}, which applies in the long-time limit \( t \gg \beta \), where the system exhibits classical diffusive behavior. In each case, we derive simplified expressions for the RMSD by appropriately approximating the general result in the respective time domain. \\
In the ballistic time domain ($t\ll\beta$), the expression of RMSD reads 
\begin{eqnarray}
(s^{2}_{reg}(t))^{per}_{\mathrm{Boson}}\approx\frac{\alpha^{\prime}}{\left(\frac{\tilde{r}_{h}}{R}\right)^2}\left(\frac{\pi}{\beta}\right)^2t^2~.
\end{eqnarray}
\noindent In the ballistic regime (\( t \ll \beta \)), the RMSD is found to grow proportionally to \( t^2 \). This result indicates that, in the early-time (ballistic) regime, the RMSD increases quadratically with time.	On the other hand in the late time domain (\(t\gg\beta\)) the expression of RMSD reads
\begin{eqnarray}
(s^{2}_{reg}(t))^{per}_{\mathrm{Boson}}\approx\frac{2\pi\alpha^\prime T}{\left(\frac{\tilde{r}_{h}}{R}\right)^2}t=D^{per}t~.
\end{eqnarray}
\begin{figure}[!h]
\begin{minipage}[t]{0.48\textwidth}
\centering\includegraphics[width=\textwidth]{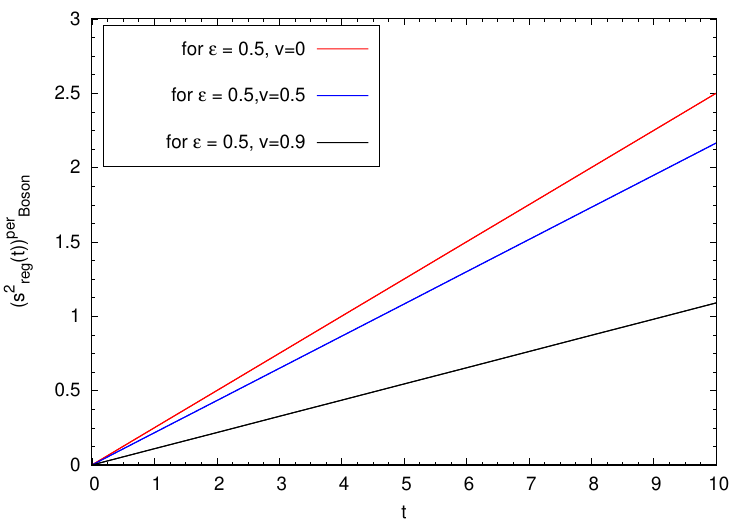}\\
\end{minipage}\hfill
\begin{minipage}[t]{0.48\textwidth}
\centering\includegraphics[width=\textwidth]{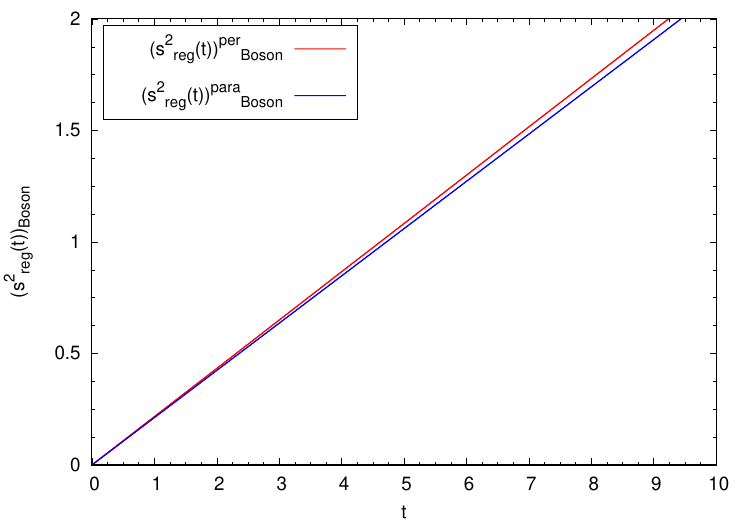}\\
\end{minipage}
\caption{The above Figures represent the variation of RMSD with respect to the observer's time. In the left panel of the above Figure we have represented the variation of RMSD with respect to the observer's time by keeping the fact that, the particle is moving perpendicular to the direction of boost. On the other hand in the right panel we have graphically represented the comparison of RMSD for parallel and perpendicular case.}
\label{fig3T}
\end{figure}
The above result indicates that, in the diffusive time regime, the  RMSD increases linearly with time. Here, \( D^{per} \) denotes the diffusion coefficient in the direction perpendicular to the applied boost. It is worth noting that the expression for the diffusion coefficient obtained in this regime matches exactly with the result previously derived in eq.\eqref{Dper}.\\
However, in the fermionic scenario,  the expression of  $(s^{2}_{reg}(t))^{per}$ reads
\begin{eqnarray}
(s^{2}_{reg}(t))^{per}_{\mathrm{Fermion}}&=&\frac{8\alpha^\prime}{\left(\frac{\tilde{r}_{h}}{R}\right)^2}\int_{0}^{\infty}\frac{d \omega}{\omega}\frac{ \sin^2(\frac{\omega t}{2})}{e^{\beta\omega}+1}=\frac{2\alpha^\prime}{\left(\frac{\tilde{r}_{h}}{R}\right)^2}\log(\frac{\left(\frac{t\pi}{2\beta}\right)}{\tanh(\frac{t\pi}{2\beta})})~.
\end{eqnarray}
In the early time domain (\(t\ll\beta\)), the expression of RMSD reduces to 
\begin{eqnarray}
(s^{2}_{reg}(t))^{per}_{\mathrm{Fermion}}\sim \frac{\pi^2\alpha^\prime}{\left(\frac{\tilde{r}_{h}}{R}\right)^2}\left(\frac{t}{\beta}\right)^2~.
\end{eqnarray}
	The above result is consistent with the corresponding expression derived in the ballistic time domain, confirming the expected short-time behavior. 
	On the other hand in the late time domain (\(t\gg\beta\)), we have
	\begin{eqnarray}\label{fper}
	(s^{2}_{reg}(t))^{per}_{\mathrm{Fermion}}\sim\frac{2\alpha^\prime}{\left(\frac{\tilde{r}_{h}}{R}\right)^2}\log(\frac{t\pi}{2\beta})~.
	\end{eqnarray}
	The above result indicates that, in the late-time regime, the system exhibits Sinai-like diffusion. This implies that, for fermions, the diffusion process becomes significantly slower compared to the bosonic case.
	\begin{figure}[!h]
		\centering
		\includegraphics[width=0.5\textwidth]{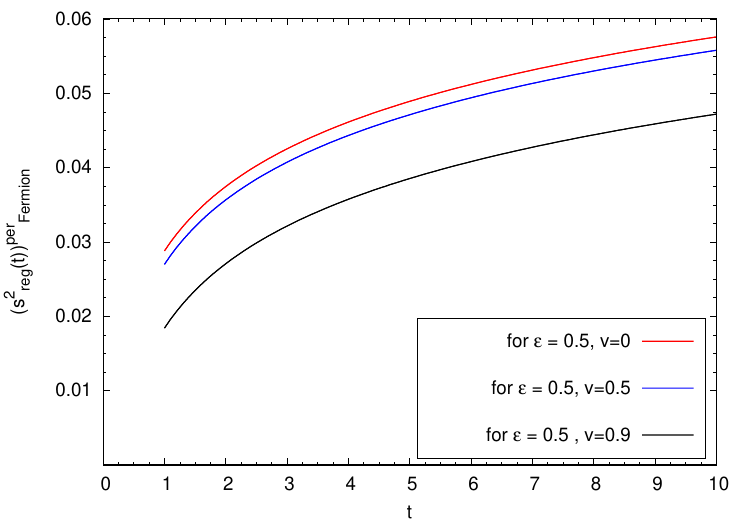}
		\caption{The Figure above shows the variation of $(s^{2}_{\mathrm{reg}}(t))^{\mathrm{per}}_{\mathrm{Fermion}}$ with respect to the observer’s time $t$. For this plot, we have considered the parameters $d = 4$, $\alpha = 10$, and $r_h = 20$. The red, blue, and black curves represent the RMSD results (as given in Eq.~\eqref{fper}) for boost velocities $v = 0$, $0.5$, and $0.9$, respectively.}
		\label{RMSDperfermion}
	\end{figure}
At this stage, we proceed to perform a comparative study of the diffusion dynamics in the parallel and perpendicular directions relative to the boost, with the aim of highlighting the anisotropic nature of the transport process. Notably, for bosons, the diffusion coefficient in the direction parallel to the boost is smaller than in the perpendicular direction, indicating that diffusion is slower along the boost direction.
\subsection{Fluctuation dissipation theorem}
\noindent In this section, we investigate the fluctuation-dissipation theorem as applied to Brownian motion in the direction perpendicular to the boost. Previously, in Section~\eqref{FDpara}, we discussed the fluctuation-dissipation theorem in the context of Brownian motion parallel to the direction of the boost. In the following, we adopt a similar methodological approach to analyze the fluctuation-dissipation theorem in the direction perpendicular to the boost. 
To proceed further let us define a symmetric Green's function as following
\begin{eqnarray}
G^{\mathrm{B,F}}_{\mathrm{Sym}}=\frac{1}{2}\left(\langle x(t)x(0)\rangle+\langle x(0)x(t)\rangle\right)~.
\end{eqnarray}
Now substituting the expressions of $\langle x(t)x(0)\rangle$ and $\langle x(0)x(t)\rangle$  (given in eq.\eqref{cor1}) in the above result, we get
\begin{eqnarray}
G^{\mathrm{B,F}}_{\mathrm{Sym}}=\frac{\alpha^\prime}{g_{xx}(r_{h})}\int_{-\infty}^{\infty}\frac{d\omega}{|\omega|}\left(\frac{2}{e^{\beta|\omega|}\pm 1}+1\right)e^{i\omega t}~.\nonumber\\
\end{eqnarray}
where \( B \) and \( F \) denote the bosonic and fermionic scenarios respectively.\\
Furthermore, in this present scenario, the fluctuation-dissipation relations for bosons and fermions can be written as follows
\begin{eqnarray}
G^{\mathrm{B,F}}_{\mathrm{Sym}}=\mathcal{F}^{-1}\left[(1+2n_{\mathrm{B,F}})\mathrm{Im}(\xi^{per}(\omega))\right]
\end{eqnarray}
where $n_{\mathrm{B,F}}$ represents the Bose-Einstein distribution (for bosons) and Fermi-Dirac distribution (for fermions). 
\noindent To compute the right-hand side of the above equation, we substitute the relevant statistical distribution function, namely, the Bose-Einstein distribution for bosons or the Fermi-Dirac distribution for fermions together with the expression for the imaginary part of the admittance provided in eq.~\eqref{adm1}. 
This yields
\begin{eqnarray}
\mathcal{F}^{-1}\left[(1+2n_{\mathrm{B,F}})\mathrm{Im}(\xi^{per}(\omega))\right]=\frac{\alpha^\prime}{g_{xx}(r_{h})}\int_{-\infty}^{\infty}\frac{d\omega}{|\omega|}\left(\frac{2}{e^{\beta|\omega|}\pm 1}+1\right)e^{i\omega t}\nonumber=G^{\mathrm{B,F}}_{\mathrm{Sym}}~.
\end{eqnarray}
This completes the verification of the fluctuation-dissipation theorem.
\section{Diffusion coefficient and butterfly velocity}
It is observed that the transport properties in a strongly correlated medium exhibits a universality, that is they show a linear resistivity over a wide range of temperature. This universality is explained by considereing a fundamental dissipative time scale $\tau\sim \frac{\hbar}{k_B T}$ \cite{Damle_1997,zaanen2004temperature}. Later in \cite{Kovtun:2004de}, it was shown that how this Planckian time scale could explain this universality, by assuming the fact that the viscosity of the strongly correlated medium control this dissipative time scale. This leads to the famous KSS bound.
Keeping this fact in mind in \cite{Hartnoll:2014lpa}, the KSS bound was reformulated in terms of the diffusion coefficient in terms of the following way
\begin{equation}
	D\sim \frac{\hbar \bar{v}^2}{k_{B}T}\sim
	\frac{h}{k_{B}}\frac{\bar{v}^2}{2\pi T}
\end{equation}
where $\bar{v}$ represents the characteristic velocity of the theory. Furthermore, for any strongly coupled theory with a dual classical gravitational background the Lyapunov exponent and butterfly velocity is related to the horizon structure of the corresponding black hole in the bulk theory \cite{Shenker:2013pqa,Shenker:2014cwa}. On the other hand it was well known from very long time that the DC transport coefficient of conserved quantities are related to the black hole horizon in the bulk theory via membrane paradigm \cite{Iqbal:2008by}. This gives  a clear idea that there is some connection between tranport phenomena and the butterfly effect.
However, in recent times in \cite{Blake:2016wvh}, this charteristic velocity is identified with the butterfly velocity of the theory. This means that the diffusion coefficient should be bounded by \cite{Blake:2016wvh}
\begin{equation}
	D\sim
	\frac{h}{k_{B}}\frac{v_B^2}{2\pi T}~.
\end{equation}
The main result of \cite{Blake:2016wvh} is that the diffusion coefficient satisfies the following relation
\begin{eqnarray}\label{ne1}
	D=C\frac{v_B^2}{2\pi T}~.
\end{eqnarray}
where $C$ is a constant and we have assume that $h=k_{B}=1$. \\
In the following subsections we will show that  the diffusion coefficients for both along and perpendicular to the boost direction can be recast in the above form systematically, and from that relation we can read off the coefficient $C$.
\subsection{Computation of the chaotic observables: Butterfly velocity $v_B$ and Lyapunov exponent $\lambda_{L}$}
\begin{figure}[!h]
		\begin{minipage}[t]{0.48\textwidth}
			\centering\includegraphics[width=\textwidth]{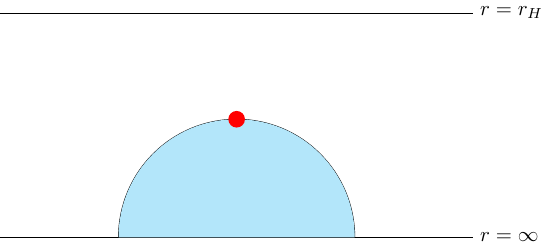}\\
		\end{minipage}\hfill
		\begin{minipage}[t]{0.48\textwidth}
			\centering\includegraphics[width=\textwidth]{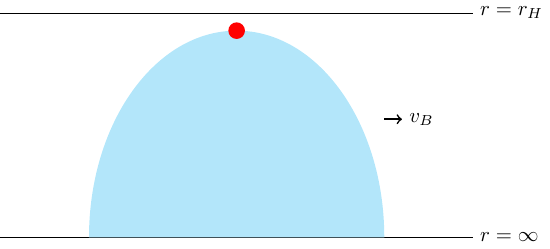}\\
		\end{minipage}
		\caption{An illustration of the entanglement wedge method of computing butterfly velocity.}
		\label{dvb}
	\end{figure}

\noindent In this section, we would provide a systematic way to derive the butterfly velocity by considering the entanglement wedge subregion duality \cite{Mezei:2016wfz,Dong:2022ucb,Mezei:2019zyt,Fischler:2018kwt} and use the obtained result to express the diffusion constant in terms of the chaotic observables. According to this duality \cite{PhysRevLett.117.021601,Wall:2012uf,Czech_2012}, a certain subregion $A$ of the boundary theory
can be completely described by a subregion in the bulk geometry. We now outline the derivation of the butterfly velocity using this method. As a first step, we analyze the boundary perspective and subsequently reinterpret it within the bulk geometry through the AdS/CFT correspondence. At the boundary a local perturbation is introduced, and the system is allowed to evolve and settle down. This process results in the scrambling of information with a constant velocity throughout the spacetime at late times. This entire boundary process can be understood through the bulk gravitational description by employing the entanglement wedge reconstruction or subregion duality. A boundary perturbation can be interpreted in the bulk as a particle falling toward the black hole from near the asymptotic boundary. It is also to be noted that the trajectory of the in falling particle lies within the extremal Ryu-Takyanagi surface. As the trajectory of the particle evolves, the Ryu–Takayanagi (RT) surface\cite{PhysRevLett.96.181602} deforms accordingly. 
At late times, the RT surface approaches the near-horizon region of the black hole with a constant velocity, known as the butterfly velocity \cite{Mezei:2016wfz}. This is represented pictorially in Fig.\eqref{dvb}. 
Some recent development in this direction can be found in \cite{Baishya:2024gih,Lilani:2024fth,Chua:2025vig,Basu:2025exh,Lilani:2025wnd}.\\
\noindent Consider perturbing the boundary via the insertion of a bulk-localized operator $\hat{V}$. This corresponds to the creation of a one-particle state in the bulk, which evolves to fall into the black hole at late times. As time time evolves, the particle in the bulk moves closer to the black hole, which implies that the $\hat{V}$ gets scrambled with an increasing number of degrees of freedom. This results in the growth of the operator in the space. It was shown that the rate of growth of this region is characterized by the butterfly velocity.
In this scenario, one can compute the butterfly velocity by using the entanglement wedge subregion duality. This duality states that a certain region in the $A$ in the boundary theory can be completely described by a sub region in the bulk. This subregion in the bulk is called the entanglement wedge ($\mathcal{M}_{A}$) of the $A$. The boundary of the entanglement wedge $\partial\mathcal{M}_{A}$ is bounded by the subsystem ($A$) and the RT surface ($\Gamma_{A}$) associated to the the region $A$. Therefore, the boundary of the entanglement wedge is defined as 
\begin{eqnarray}
\partial\mathcal{M}_{A}=A\cup\Gamma_{A}~.
\end{eqnarray}
Let us consider a generic black hole spacetime of the following form
\begin{eqnarray}\label{vb1}
ds^2=-G_{tt}(r)~dt^2+G_{rr}(r)~dr^2+G_{ij}(r)dx^{i}dx^{j}
\end{eqnarray}
where $r$ is the bulk coordinate and ($t,x^{i}$) represent boundary coordinate and $ i,j=1,...,(d-1)$. Further, the boundary is located at $r=\infty$ and the horizon is at $r=r_{H}$. To proceed further, let us consider a constant time slice of the geometry, taken at a sufficiently late time following the application of $\hat{V}$, such that the operator has effectively delocalized across a large boundary region $A$. This simplifies our analysis. This consideration results in a linearized equation of motion for the entanglement wedge because the RT surface corresponding to the region $A$ lies near the black hole horizon. Therefore, in the near horizon limit the spacetime metric given in eq.\eqref{vb1} can be expressed in terms of Rindler coordinates. In the near horizon approximation, the metric coefficients can be written as 
\begin{eqnarray}
G_{tt}&=&c_{0}(r-r_{H})~~;~~G_{rr}=\frac{c_1}{(r-r_{H})}\nonumber\\G_{ij}(r)&=&G_{ij}(r_{H})+(r-r_{H})~G^{\prime}_{ij}(r_{H})
\end{eqnarray}
where the coefficients $c_{1}$ and $c_{2}$ are related to the inverse of Hawking temperature in the following way
\begin{eqnarray}
\beta=4\pi\sqrt{\frac{c_{0}}{c_{1}}}~.
\end{eqnarray}
In order to facilitate the analysis, we now perform the following coordinate transformation
\begin{eqnarray}
(r-r_{H})=\frac{\rho^2}{G^{\prime}_{tt}(r_{H})}\left(\frac{2\pi}{\beta}\right)^2~.
\end{eqnarray}
Thus, in the near-horizon limit the spacetime metric (given in eq.~\eqref{vb1}) in terms of the new coordinate $\rho$ can be written as
\begin{eqnarray}\label{vb2}
ds^2=-\left(\frac{2\pi}{\beta}\right)^2\rho^2 dt^2+ d\rho^2+\left(G_{ij}(r_{H})+\frac{G^{\prime}_{ij}(r_{H})}{G^{\prime}_{tt}(r_{H})}\left(\frac{2\pi}{\beta}\right)^2\rho^2\right)dx^{i}dx^{j}~.
\end{eqnarray}

Now we calculate the area of the RT surface which defines the entanglement wedge of the boundary region $A$. To do this we first consider a constant time slice and then parametrize the bulk coordinate $\rho$ in terms of the boundary coordinates $x^{i}$, that is, $\rho=\rho(x_i)$. Keeping this in mind, the induced metric on the constant time slice can be written as
\begin{eqnarray}
ds^{2}|_{\mathrm{ind}}=\left[(\partial_{i}\rho)^2+G_{ii}(r_{H})+\frac{G^{\prime}_{ii}(r_{H})}{G^{\prime}_{tt}(r_{H})}\left(\frac{2\pi}{\beta}\right)^2\rho^2\right](dx^{i})^2~.
\end{eqnarray}
Therefore, the area functional can be written as
\begin{eqnarray}
\mathrm{Area}=\sqrt{\det(G_{ii}(r_{H}))}\int d^{d-1}x\left[1+\frac{G^{\prime}_{ii}(r_{H})}{G^{\prime}_{tt}(r_{H})G_{ii}(r_{H})}\left(\frac{2\pi}{\beta}\right)^2\rho^2+\frac{1}{G_{ii}(r_{H})}(\partial_{i}\rho)^2\right]^{\frac{1}{2}}~.
\end{eqnarray}
In the limit $\rho<<1$ the area functional can be written as
\begin{eqnarray}
\mathrm{Area}\approx\frac{\sqrt{\det(G_{ii}(r_{H}))}}{2}\int d^{d-1}x\left[\frac{G^{\prime}_{ii}(r_{H})}{G^{\prime}_{tt}(r_{H})G_{ii}(r_{H})}\left(\frac{2\pi}{\beta}\right)^2\rho^2+\frac{1}{G_{ii}(r_{H})}(\partial_{i}\rho)^2\right]~.
\end{eqnarray}
Now one can obtain the equation of motion of the entanglement wedge by varying the above area functional with respect to $\rho$. This results
\begin{eqnarray}\label{vb3}
\frac{\partial_{i}^2\rho}{G_{ii}(r_{H})}=\mu^2\rho~~,~~\mu^2=\frac{G^{\prime}_{ii}(r_{H})}{G^{\prime}_{tt}(r_{H})G_{ii}(r_{H})}\left(\frac{2\pi}{\beta}\right)^2~.\nonumber\\
\end{eqnarray}
To solve the above equation, we introduce a new coordinate $\sigma^i$, defined by
\begin{eqnarray}
\sigma^{i}=x^{i}\sqrt{G_{ii}(r_{H})}~.
\end{eqnarray}
In this new coordinate system, the differential equation given in Eq.~\eqref{vb3} can be rewritten as follows
\begin{eqnarray}
\left(\frac{\partial}{\partial \sigma^{i}}\right)^2\rho=\mu^2\rho~.
\end{eqnarray}
The solution of the above equation can be written as \cite{Mezei:2016wfz,Liu:2013una}
\begin{eqnarray}
\rho(\sigma^{i})=\rho_{min}\frac{\Gamma(a+1)}{2^{-2}\mu^a}\frac{I_{a}(\mu|\sigma|)}{|\sigma|^a}~~,~~ a=\frac{d-3}{2}~.
\end{eqnarray}
In this above result, $\rho_{min}$ denotes the distance of closest approach to the black hole horizon and $I_{a}$ is the Bessel function of second kind. Now if $\rho$ exceeds $\beta$, the surface lies near the horizon and reaches the boundary very quickly. Thus, we can determine the size of the region $A$ by solving the following differential equation
\begin{eqnarray}\label{New1}
\beta=\rho_{min}\frac{\Gamma(a+1)}{2^{-2}\mu^a}\frac{I_{a}(\mu|R_\sigma|)}{|R_\sigma|^a}
\end{eqnarray}
where $R_{\sigma}$ represents the size of the region $A$ in $\sigma$ coordinate. One can solve the above equation in large $R_{\sigma}$ limit. In this limit the Bessel function can be approxiamted as 
\begin{eqnarray}
I_{a}(\mu|R_\sigma|)\sim \frac{e^{\mu R_{\sigma}}}{\sqrt{2\pi \mu R_{\sigma}}}~.
\end{eqnarray}
Therefore, using the above result in eq.\eqref{New1} we get the following expression of $\rho_{min}$ \cite{Fischler:2018kwt}
\begin{eqnarray}
\rho_{min}\approx e^{-\mu R_{\sigma}}~.
\end{eqnarray}
Furthermore, let us assume that, $R_{i}$ represents the size of the region $A$ along $x^{i}$ direction. Now in the $\sigma$ coordinates this size of the region is described by $R_{\sigma i}=\sqrt{G_{ii}(r_{H})}R_{i}$. Now we consider the scenario where the infalling particle created by $\hat{V}$ lies entirely within the entanglement wedge. This condition leads to the following constraint
\begin{eqnarray}\label{New2}
\rho_{min}\le \rho(t)~.
\end{eqnarray}
On the other hand as the particle gets blue shifted as it goes closer to the black hole horizon. This indicates that the particle asymptotically approaches the horizon. One can obtain the radial trajactory of this infalling particle by studying the null geodesic in the background metric given in eq.\eqref{vb2}. Therefore the equation of motion for the null infalling geodesic reads \footnote{We obtain this equation of motion by setting $ds^2=0$ in eq.\eqref{vb2} and fixing all other spatial corrdinates.}
\begin{eqnarray}
\frac{d\rho}{dt}=-\frac{2\pi}{\beta}\rho
\end{eqnarray}
the negative sign indates that we are considering the infalling motion. Thus the solution of the above equation can be written as 
\begin{eqnarray}\label{vb4}
\rho(t)=\rho_{0}e^{-\frac{2\pi}{\beta}t}~.
\end{eqnarray}
Now keeping this result in mind and using the expression of $\rho_{min}$ in eq.\eqref{New2} we get
\begin{eqnarray}
R_{i}\ge v_{B,x_{i}}t
\end{eqnarray}
where $v_{B,x_{i}}$ denotes the butterfly velocity along the $x_{i}$ direction \footnote{Here we have chosen $\rho_{0}=1$}, which is given by the following expression
\begin{eqnarray}\label{VB}
v_{B,x_{i}}&=&\frac{2\pi}{\beta}\frac{1}{\mu\sqrt{G_{ii}(r_{H})}}
=\sqrt{\frac{G^{\prime}_{tt}(r_{H})}{G_{ii}(r_{H})}}\frac{1}{\sqrt{\frac{G^{\prime}_{kk}(r_{H})}{G_{kk}(r_{H})}}}~.
\end{eqnarray}
Before finishing this section, we would like to mention that, this general expression of butterfly velocity obtained above by using the entanglement wedge subregion duality matches exactly with that obtained using the shock-wave method.\\
Furthermore, as the boosted black brane is a maximally chaotic systems the Lyapunov exponent can be obtained by saturating the MSS bound \cite{Maldacena:2015waa}. This reads 
\begin{eqnarray}\label{lambpara}
\lambda^{v}_{L}=\frac{2\pi}{\beta}=2\pi T
\end{eqnarray}
where $T$ is the Hawking temperature. Now substituting the expression of Hawking temperature (given in eq.\eqref{T}) in the above result, we get the following expression of the Lyapunov exponent to be
\begin{eqnarray}\label{Ly}
\lambda^{v}_{L}=\frac{d}{2\pi R^2}\frac{r_{H}}{\gamma}~.
\end{eqnarray}
\subsection{Spatial diffusion constant in terms of $v_B$ and $\lambda_{L}$, parallel the boost}
In the previous section, we reviewed the systematic procedure for computing the butterfly velocity by employing the entanglement wedge subregion duality for a generic black hole metric. In this section, we apply this method to compute the butterfly velocity in a boosted thermal plasma. Specifically, we consider the case where the operator $\hat{V}$ is applied along the direction of the boost, namely the $y$-direction. It is also to be noted that as we interested in the leading order correction in butterfly velocity due to boost. Therefore, we consider to proceed by considering the constant time slice approximation.
Following the approach outlined above, the butterfly velocity in this scenario can be computed by substituting the metric components (given in Eq.\eqref{matricc}) into the general expression provided in Eq.\eqref{VB}. This yields
\begin{eqnarray}\label{vbpara}
(v_{B}^2)^{para}=\frac{1-\gamma^2\left(1-\frac{d}{2}\right)}{(d-1)+\frac{\gamma^2v^2 (d-2)}{2}}~.
\end{eqnarray}
We note that, in the limit $v \rightarrow 0$, the result smoothly reduces to the corresponding expression for the Schwarzschild-AdS black hole in $(d+1)$ spacetime dimension.\\
\noindent It is well known in the literature \cite{Maldacena:2015waa,Blake:2016wvh} that the diffusion coefficient $(D^{para})$ is related to the Lyapunov exponent $(\lambda_{L})$ and butterfly velocity $(v_B)$ in the following way 
\begin{eqnarray}
D^{para}=C^{para}\frac{(v_{B}^2)^{\mathrm{para}}}{\lambda_{L}^{v}}
\end{eqnarray}
where $C^{para}$ is a constant.  We can determine this constant $C^{para}$ in the following way.\\ The diffusion coefficient for the parallel case can be written as 
\begin{eqnarray}\label{Ne1}
D=\Sigma(\tilde{r}_{h}) T=\Sigma(\tilde{r}_{h})\frac{d r_{h}}{4\pi R^2\gamma}
\end{eqnarray}
where $\Sigma(\tilde{r}_{h})$ is given as 
\begin{equation*}
	\Sigma(\tilde{r}_{h})=\frac{2\pi\alpha^\prime }{\left(\frac{\tilde{r}_h}{R}\right)^2\left(1+v^2\gamma^2\left(\frac{r_{h}}{\tilde{r}_{h}}\right)^d\right)^{\frac{1}{2}}}\frac{\left[1-\gamma^2\left(\frac{r_{h}}{\tilde{r}_{h}}\right)^d\right]^{\frac{1}{2}}}{f^{\frac{1}{2}}(\tilde{r}_{h})}~.
\end{equation*}
Now  from the expression of butterfly velocity given in eq.\eqref{vbpara}, one can substitute the expression of $d$ in the following form
\begin{eqnarray}
	d=\frac{2}{\gamma^2}\left[(v_{B}^2)^{para}\left((d-1)+\frac{v^2\gamma^2(d-2)}{2}\right)+\gamma^2-1\right]~.
\end{eqnarray}
Now substituting the above expression of $d$ in the results of diffusion coefficient in eq.\eqref{Ne1}, we get 
\begin{eqnarray}
	D=2\Sigma(\tilde{r}_{h})\frac{r_{h}}{4\pi R^2\gamma}\frac{2}{\gamma^2}\left[(v_{B}^2)^{para}\left((d-1)+\frac{v^2\gamma^2(d-2)}{2}\right)+\gamma^2-1\right]~.
\end{eqnarray}
Now using the expression of Hawking temperature, we substitute  $\frac{r_{h}}{4\pi R^2\gamma}=\frac{T}{d}$ in the above. This results
\begin{eqnarray}
	D=\frac{4\pi \Sigma(\tilde{r}_{h})T^2\left[(d-1)+\frac{v^2\gamma^2(d-2)}{2}\right]}{d\gamma^2}\frac{(v_{B}^2)^{para}}{2\pi T}+ \frac{2\Sigma(r_{h})T (\gamma^2-1)}{\gamma^2 d}~.
\end{eqnarray}
Further, the second term in the above result  can be expressed in terms of the diffusion coefficient (from eq.\eqref{Ne1}) and after further simplification we can rewrite the above result as
\begin{eqnarray}
	D=\frac{4\pi \Sigma(\tilde{r}_{h})T^2\left[(d-1)+\frac{v^2\gamma^2(d-2)}{2}\right]}{d\gamma^2\left[1-\frac{2(\gamma^2-1)}{d\gamma^{2}}\right]}\frac{(v_{B}^2)^{para}}{2\pi T}~.
\end{eqnarray}
The above result of diffusion coefficient exactly has the similar form as given in eq.\eqref{ne1}. Therefore, we can read off the above constant which has the following form
\begin{eqnarray} 
C^{\mathrm{para}}=\frac{4\pi \Sigma(\tilde{r}_{h})T^2\left[(d-1)+\frac{v^2\gamma^2(d-2)}{2}\right]}{\left[1-\frac{2(\gamma^2-1)}{d\gamma^{2}}\right]\gamma^2 d}~.
\end{eqnarray}
The above relation is valid for any value for the boost. It is to be noted that the value of the constant $C^{\mathrm{para}}$ is less in the presence of the boost ($v\ne 0$). This tells us that the ratio of the $D^{\mathrm{para}}$ and $((v_{B}^2)^{\mathrm{para}}/\lambda_{L}^v)$ is less in the presence of the boost compared to the value in the absence of the boost. 

\subsection{Spatial diffusion constant in terms of $v_B$ and $\lambda_{L}$, perpendicular to the boost}
In this section, we again compute the butterfly velocity, this time by perturbing a region that is perpendicular to the direction of the boost. In this case we have applied the operator $\hat{V}$ along the $x$. Therefore, to compute the butterfly velocity in this scenario we have to substitute the metric components in the general expression $v_{B}$ given in eq.\eqref{VB}. This results in
\begin{eqnarray}\label{vbper}
(v_{B}^2)^{per}=\frac{\left(1-\gamma^2\left(1-\frac{d}{2}\right)\right)\left(1+v^2\gamma^2\right)}{(d-1)-\gamma^2v^2}~.
\end{eqnarray}
On the other hand the expression of the Lyapunov exponent is given by eq.\eqref{Ly}. \\
Now following the similar procedure as explained in the earlier section we can express the result of diffusion coefficient perpendicular to the direction of boost in following form
\begin{eqnarray}
D^{\mathrm{per}}=C^{\mathrm{per}}\frac{(v_{B}^2)^{\mathrm{per}}}{\lambda_{L}^{v}}
\end{eqnarray}
where the expression for the proportionality constant $C^{\mathrm{per}}$ reads
\begin{eqnarray}
C^{\mathrm{per}}=\frac{8\pi \alpha^\prime T^2\left[(d-1)-v^2\gamma^2\right]}{\left(\frac{\tilde{r}_h}{R}\right)^2\left[1-\frac{2(\gamma^2-1)}{d\gamma^{2}}\right]\gamma^2 d\left(1+v^2\gamma^2\right)}~.
\end{eqnarray}
\section{Conclusion}
\noindent We now represent a summary of our work. In this work we have studied the Brownian motion of a heavy particle in a thermal plasma which is moving along a particular direction in the boundary field theory holographically. To do this analysis we have considered boosted AdS Schwarzschild black hole in the bulk. This black hole geometry in the bulk represents thermal plasma moving along a particular direction in the boundary field theory. We have studied the Brownian motion of this massive particle both along the boost and perpendicular to the boost. One can studied this problem by considering the fluctuation of a string suspended from the horizon in both the parallel and perpendicular directions relative to the direction of the boost. In our work we have computed the diffusion coefficient in both the scenarios by two different approaches. We have shown that the diffusion coefficient obtained in both the scenarios by two different methods matches exactly. We have also studied the fluctuation dissipation theorem in both the cases. Our analysis indicates that the presence of a boost leads to a suppression of the diffusion process in both the parallel and perpendicular directions. Furthermore, our findings indicate that Brownian motion parallel to the boost experiences a slower diffusion process than motion in the perpendicular direction. We have also represented this observation graphically. In essence, the direction-dependent redshift, anisotropic thermal noise, and momentum dissipation rates introduced by the boost are responsible for the suppression of diffusion, especially in the parallel direction. This observation hold only for bosons. In contrast, our analysis reveals that fermions exhibit Sinai-like diffusion, characterized by extremely slow, logarithmic-in-time spreading of the probability distribution, which is markedly different from the behavior observed in the bosonic case. This observation holds for both the parallel and perpendicular cases.\\
First we consider the Brownian motion of a heavy particle in thermal boosted plasma along the direction of the boost. To do this study we have considered the fluctuation of an open string along the direction of the boost in the boosted AdS-Schwarzschild background. We have assumed that one end point of this open string is attached with the black hole horizon and the other end is free to move. Then we have studied the fluctuation of the string end point which is free to move along the direction of the boost by considering the Nambo-Goto action. It is worth noting that previous studies have been conducted in a generic diagonal spacetime background. However, in the present work, we consider a non-diagonal spacetime metric, which introduces additional complexities and renders the analysis non-trivial. Then extremizing the NG action we have obtained the equation of motion of the string fluctuation. However, it is impossible to get an exact solution of this differential equation. It is also worth noting that an exact solution to the string fluctuation can be obtained only in the BTZ black hole background in 
($2+1$)-dimensional spacetime. Therefore, to get an approximated solution we have used the standard patching method. This method allows us to obtain the string fluctuation solutions across different regimes of the theory. First we consider the near horizon region, that is, $r\sim r_h$, which is also known as the IR region. To get the solution in this regime we have to recast the equation of motion in terms of the tortoise coordinates. This simplifies the calculation a lot. It should be kept in mind that, we are in the low frequency domain. Then we have moved on to get the solution of the string fluctuation in the hydrodynamic limit, by considering the fact that, $\omega \rightarrow 0$. Then comparing these two solutions, we have fixed the constants appearing in solving the differential equation. Then we proceed to obtain the solution in the UV region, $r\rightarrow \infty$, by using the solution in the hydrodynamic limit. Keeping these results in mind, we have computed the diffusion coefficient by two completely different methods. First, we compute the diffusion coefficient by  computing the admittance. To do this we have applied an external electric field at the boundary. This modifies the boundary condition of the string dynamics. It does not effect the bulk dynamics of the string. Then we proceed to compute the admittance by considering the solution of string fluctuation in the UV region. Keeping this result of admittance in mind, we  compute the diffusion coefficient. We have observed that the diffusion coefficient is modified due to the effect of the boost. We have found that the diffusion process get slower due to the effect of the boost. Then we proceed to compute the diffusion coefficient by computing the regularized mean squared displacement. It can be done considering the solution of the string fluctuation in the UV region. This solution near the boundary ($r\sim r_{b}$) can be identified as the displacement of the heavy quark at the boundary. We have computed a general expression of $s^{2}_{reg}(t)$ for both fermions and bosons. We have observed that in the early time domain (also known as the ballistic time domain), the regularized mean squared displacement grows quadratically with observer's time, that is, $s^2_{reg}(t)\sim t^2$ for both the bosons and fermions. On the other hand in the late time domain (in the diffusive regime), we have shown that, the RMSD has different form for bosons and fermions. In particular, for bosons we have shown that RMSD increases linearly with time, that is, $s^2_{reg}(t)\sim t$. The slope of the $s^2_{reg}(t)$ vs $t$ curve is nothing but the diffusion coefficient. Thus, we obtain the expression diffusion coefficient for bosons. These findings are illustrated graphically. The analysis shows that the diffusion coefficient diminishes as the boost increases. This behavior is reflected in the \( s^2_{\text{reg}}(t) \) versus \( t \) plots, where the slope of the curve, which corresponds to the rate of diffusion becomes progressively smaller with increasing boost. On the other hand, for fermions, we have found that the regularized mean squared displacement behaves as $s^2_{reg}(t) \sim \log(t)$, indicating a sub-diffusive regime. This type of ultra-slow diffusion is characteristic of Sinai diffusion, as observed in ultra-slow scaled Brownian motion. Then we moved on to check the fluctuation dissipation theorem for both the bosons and fermions in this context. To do this, we have first constructed the symmetric Green's function $G_{\mathrm{Sym}}$. Then we have shown that this symmetric Green's function is related to the imaginary part of admittance via the relation $	G^{\mathrm{B,F}}_{\mathrm{Sym}}=\mathcal{F}^{-1}\left[(1+2n_{\mathrm{B,F}})\mathrm{Im}(\xi(\omega))\right]$. This completes the check of the fluctuation-dissipation theorem.\\
A similar analysis has also been carried out for Brownian motion perpendicular to the direction of the boost, following the approach described above. We have observed that in this case also the diffusion coefficient decreases with the boost. Notably, for a fixed boost parameter, the diffusion coefficient along the direction of the boost is consistently lower than that in the transverse direction, that is, $D^{para}<D^{per}$. This observation implies that diffusion is considerably slower along the direction of the boost compared to the perpendicular direction. Similar to the parallel case, here we have also computed the RMSD for both bosons and fermions. We have also shown that in the ballistic time domain $s^{2}_{reg}(t)\sim t^2$ for both fermions and bosons. On the other hand, in the late time domain we have shown $s^{2}_{reg}\sim t$ for bosons. The slope of the this $s^{2}_{reg}$ versus $t$ curve represents the diffusion coefficient. Again we have observed that the slope of this curve decreases with the boost parameter, which indicates the fact that the diffusion coefficient decreases with the increment in the boost parameter. We have also graphically compared the results of the regularized mean squared displacement (RSMD) for bosons in both the parallel and perpendicular directions with respect to the boost. The comparative analysis reinforces our previous result, providing additional evidence for the directional dependence of the diffusion process, that is,  $D^{para}<D^{per}$. On the other hand, for fermions it is found that, $s^2_{reg} (t)\sim \log(t)$. This again indicates ultra slow Brownian motion known as the  Sinai diffusion. We have also checked the fluctuation-dissipation theorem in this scenario by following the similar approach. Finally, we make use of the entanglement wedge subregion duality to holographically compute the butterfly velocity. We have then systematically expressed the diffusion coefficients in terms of the chaotic observables, that are, Lyapunov exponent and butterfly velocity. In doing so, the explicit expression for the proportionality constants have been obtained. This procedure in turn provides us a subtle realization about the chaotic origin of the diffusive properties. 
\section{Acknowledgement}
\noindent ARC would like to thank SNBNCBS for the Bridge fellowship.
\section{Appendix: Brief discussion on linear response theory and fluctuation-dissipation theorem}
\noindent In this section, we briefly discuss the \textit{linear response theory} and the \textit{fluctuation-dissipation theorem} \cite{RKubo_1966}. We begin with the linear response theory.
Consider a system that initially evolves under the Hamiltonian \( H_0 \). Suppose the system is then subjected to a time-dependent external perturbation. As a result, the total Hamiltonian of the system after the perturbation can be written as
\begin{equation}
H(t) = H_0 - \lambda A(\Gamma) f(t)
\end{equation}
where \( A \) is the phase space observable ($\Gamma$ depends on the coordinates and momenta) of the system that couples to the external field, and \( f(t) \) represents the time-dependent perturbing field and \(\lambda\) represents the strength of the perturbation. The response of the system can be measured by computing the expectation value of any observable \( B(\Gamma) \) in the presence of the perturbation. Let $\rho_{0}$ be the density matrix describing the system. Therefore, in the absence of any perturbation, the expectation value of a quantity $B(\Gamma)$ can be written as
\begin{eqnarray}\label{xB0}
\langle B(\Gamma)\rangle_{\rho_{0}}=\int d\Gamma \rho_{0}(\Gamma) B(\Gamma)=\int d\Gamma \frac{e^{-\beta H_{0}}}{Z} B(\Gamma)=B_{0}\nonumber\\
\end{eqnarray}
where $Z$ is the partition function in the absence of any perturbation and $\rho_{0}(\Gamma)=\frac{e^{-\beta H_{0}}}{Z}$. On the other hand in the presence of the perturbation $H^{\prime}(t)$, the expectation value of $B(\Gamma)$ can be written as
\begin{eqnarray}\label{xB}
\langle B(\Gamma)\rangle_{\rho(t)}=\int d\Gamma \rho(\Gamma,t) B(\Gamma)
\end{eqnarray}
where $\rho(\Gamma,t)$ is not the distribution in thermal equilibrium.\\
The time evolution of the density matrix $\rho(\Gamma,t)=\rho(t)$ is given by the following differential equation
\begin{eqnarray}\label{t1}
\frac{d \rho(t)}{dt}=-\{H(t),\rho(t)\},
\end{eqnarray}
where $\{.,.\}$ represents the Poisson bracket. 
To proceed further, we take the following ansatz for $\rho(t)$
\begin{eqnarray}
\rho(t)=\rho_{0}+\lambda\Delta\rho(t)~.
\end{eqnarray}
The above form of $\rho(t)$ has been considered to facilitate the analysis of the system's response to perturbations, specifically focusing on effects that are linear in the perturbation strength. It is also to be noted that $\rho_{0}$ obeys $\frac{d\rho_{0}}{dt}=0,\{H_{0},\rho_{0}\}=0$. Keeping this in mind and substituting the above expression of $\rho(t)$ in eq.\eqref{t1}, we get
\begin{eqnarray}
\frac{d \Delta \rho(t)}{dt} = i \mathcal{L}_0 \Delta \rho(t) + f(t) \{A, \rho_0\} + \mathcal{O}(\lambda), \quad \Delta \rho(-\infty) = 0\nonumber\\
\end{eqnarray}
where $\mathcal{L}_0 =i\{H_{0},\Delta\rho(t)\}$. The formal solution of the above equation can be written as 
\begin{eqnarray}
\Delta\rho(t)=\int_{-\infty}^{t}ds~e^{i\mathcal{L}_{0}(t-s)}\{A,\rho_{0}\}f(s)~.
\end{eqnarray}
Now one can compute the expectation value of an observable in the presence of perturbation by substituting the above result in eq.\eqref{xB}. This yields
\begin{eqnarray}
\langle B\rangle_{\rho(t)}=B_{0}+\lambda\Delta B(t)
\end{eqnarray} 
where $B_{0}$ is the expectation value of the observable $B$ in the absence of perturbation (given by eq.\eqref{xB0}) and $\Delta B(t)$ represents the change in $B$ due to perturbation. The expression of $\Delta B(t)$ is given by 
\begin{eqnarray}\label{DelB}
\Delta B(t)&=&\int_{-\infty}^{t} ds~f(s)\int d\Gamma e^{i\mathcal{L}_{0}(t-s)}\{A,\rho_{0}\}B\nonumber\\&=&\int_{-\infty}^{t} ds \, f(s) \, \chi(t - s)
\end{eqnarray}
where $\chi(t)$ is the response function which is given by
\begin{eqnarray}
\chi(t)=\int d\Gamma e^{i\mathcal{L}_{0}t}\{A,\rho_{0}\}B~.
\end{eqnarray}
It is to be noted that the response function is independent of the external time dependent field. One can also write the response function in the following form
\begin{eqnarray}
\chi(t)=-\langle\{A(t),B\}\rangle_{\rho_{0}}~.
\end{eqnarray}
Furthermore, substituting the expression of $\rho_{0}=e^{-\beta H_{0}}/Z$ in the expression of the response function, we have
\begin{eqnarray}
\chi(t)=\beta\langle\dot{A(t)}B\rangle_{\rho_{0}}~.
\end{eqnarray}
This completes the analysis of the linear response theory.\\ Now we proceed to discuss the fluctuation-dissipation theorem. This theorem relates the time dependent correlation function $C_{AB}(t)$ of two observables $A$ and $B$ to the response function $\chi_{AB}$. To proof this theorem we proceed as follows.\\
We assume that the perturbing field \( f(t) \) takes the form of a plane wave and is introduced into the system adiabatically. Thus, the perturbing field has the following form
\begin{eqnarray}
f(t)=f_{0}e^{-i(\omega+i\epsilon)t}~~;~~\epsilon\rightarrow 0~.
\end{eqnarray}
Now using the above form of the perturbing field in eq.\eqref{DelB}, the change in the expectation value of $B$ can be written as
\begin{eqnarray}
\Delta B(t)=\chi(\omega)f_{0}e^{-i\omega t}~
\end{eqnarray}
where $\chi(\omega)$ is the Laplace transform of $\chi(t)$, which is given as 
\begin{eqnarray}
\chi(\omega)=\lim_{\epsilon\rightarrow 0^{+}}\int_{0}^{\infty}\chi(t)~e^{i(\omega+i\epsilon)t}\equiv\chi^{\prime}(\omega)+i\chi^{\prime\prime}(\omega)~.\nonumber\\
\end{eqnarray}
The Laplace transform can be performed by keeping \( \epsilon \) finite, which ensures the convergence of the integral. This helps us to get the analytical continuation of $\chi(\omega)$ to the complex frequency $z=\omega+i\epsilon$. This yields
\begin{eqnarray}
\chi(z)&=&\beta\langle AB\rangle_{\rho_{0}}-iz\beta\int_{0}^{\infty} dt~\langle A(t)B\rangle_{\rho_{0}}e^{izt}\nonumber\\&=&\beta\langle AB\rangle_{\rho_{0}}-iz\beta\tilde{C}_{AB}(z)~.
\end{eqnarray}
The Laplace transform of the correlation function can be obtained from its Fourier transform by analytic continuation. This results in getting
\begin{eqnarray}
\tilde{C}_{AB}(z)=i\int_{-\infty}^{\infty}d\omega^{\prime} \frac{C_{AB}(\omega^{\prime})}{z-\omega^{\prime}}~.
\end{eqnarray}
The real part of $\tilde{C}_{AB}(z)$ reads 
\begin{eqnarray}
\mathcal{R}e(\tilde{C}_{AB}(z))=\pi C_{AB}(\omega)~.
\end{eqnarray}
Finally, the imaginary part of the response $\chi(z)$ reads
\begin{eqnarray}
\mathcal{I}m(\chi(z))=\chi^{\prime\prime}(\omega)=\beta \pi\omega C_{AB}(\omega).
\end{eqnarray}
In the above result, we have set $\epsilon=0$. The above relation is nothing but the fluctuation-dissipation theorem. The above equation can also be written as 
\begin{eqnarray}
\chi^{\prime\prime}(\omega)=\frac{\pi\omega}{T} C_{AB}(\omega)~.
\end{eqnarray}
The left hand side of the above equation represents dissipation, on the other hand the RHS denotes the fluctuation. This is the classical version of the fluctuation-dissipation theorem.
\section{Appendix II : Tortoise coordinate for backgrounds with off-diagonal metric components}
In this appendix, we will obtain the expression of the tortoise coordinate for a general stationary background containing a mixed $t-y$ component in the metric. Consider the bulk geometry of the form
\begin{eqnarray}
ds^2&=&-|g_{tt}(r)|\,dt^2+g_{rr}(r)\,dr^2+2g_{ty}(r)\,dt\,dy+g_{yy}(r)\,dy^2\nonumber\\&+&g_{ij}(r)\,dx^{i}dx^{j},
\end{eqnarray}
where all metric components depend only on the radial coordinate $r$. The presence of the off-diagonal term $g_{ty}$ implies that the time translation Killing vector $\partial_t$ does not generate the event horizon. Instead, the horizon is generated by a linear combination of Killing vectors. Therefore, the horizon generating Killing vector can be written as
\begin{equation}
\chi=\partial_t+\Omega_H \partial_y ,
\end{equation}
where $\Omega_H$ denotes the effective boost (or angular velocity) associated with the horizon. The location of the event horizon $r=r_H$ is determined from the condition that $\chi$ becomes null,
\begin{equation}
\chi^2=g_{tt}+2\Omega_H g_{ty}+\Omega_H^2 g_{yy}=0 \quad \text{at} \quad r=r_H .
\end{equation}
Now substituting the expressions of the metric components in the above result we get 
\begin{eqnarray}
\Omega_H=-v~.
\end{eqnarray}
Now one can diagonalise the above metric by introducing a new coordinate $\tilde{t}$ as follows
\begin{equation}
d\tilde{t}=dt-v dy.
\end{equation}
In this coordinate system, the $(\tilde t,r)$ sector of the metric takes the form
\begin{equation}
ds^2 \simeq -F(r)\, d\tilde t^{\,2} + g_{rr}(r)\, dr^2 + \cdots ,
\end{equation}
where the effective redshift function is defined as
\begin{equation}
-F(r)= g_{tt}-2v g_{ty}+v^2 g_{yy}.
\end{equation}
The tortoise coordinate $r_*$ is introduced such that radial null trajectories satisfy
\begin{equation}
d\tilde t = \pm dr_* .
\end{equation}
This requirement leads to the definition
\begin{equation}
\frac{dr_*}{dr}=\sqrt{\frac{g_{rr}(r)}{|F(r)|}}.
\end{equation}
Consequently, the tortoise coordinate can be written as
\begin{equation}
r_*=\int dr \sqrt{\frac{g_{rr}(r)}{\left(g_{tt}-2v g_{ty}+v^2 g_{yy}\right)}}.
\end{equation}
Now again substituting the explicit form of the metric components, leads to the following expression of the tortoise coordinate
\begin{equation}
r_{*}=\int\frac{\gamma R^2 dr}{r^2 f(r)}~.
\end{equation}
\bibliographystyle{hephys}  
\bibliography{Reference}
\end{document}